\documentclass{aa}  

\usepackage{graphicx}

\usepackage{txfonts}
\usepackage{hyperref} % Enables hyperlinks

\begin{document}

   \title{ Search for High Energy  Neutrino Sources using IceCube Neutrino Events and \textit{Fermi}/LAT Observations}
   \titlerunning{Search for High Energy Neutrino Sources}

   \author{R. Neupane\inst{1,2}
          \and
          N. Dhital\inst{1}\thanks{Corresponding author: {niraj.dhital@cdp.tu.edu.np}}
          }

\institute{
    Central Department of Physics, Tribhuvan University, Kirtipur, Kathmandu, Nepal
%    \email{rajendra.785711@iost.tu.edu.np}, \email{niraj.dhital@cdp.tu.edu.np}
    \and
    Department of Physics, Birendra Multiple Campus, Tribhuvan University, Bharatpur, Nepal
}

   \date{Received xxxxxx xx, xxxx}

% \abstract{}{}{}{}{} 
% 5 {} token are mandatory
 
  \abstract
   {The sources of high-energy astrophysical neutrinos remain largely unknown. Multi-wavelength observational campaigns have identified a few plausible candidates, but their numbers are limited.}
   {We search for an excess of high-energy neutrino events originating from the directions of blazar sources, compared to the expected background. Additionally, we investigate blazars located within a small angular distance from the arrival directions of high-energy neutrinos, analyzing their characteristics to assess their plausibility as neutrino sources.}
   {We generate random samples of events following the right ascension and the declination distributions of the high-energy neutrino events detected by the IceCube Neutrino Observatory and count the coincidences of generated samples as well as the true observations with the $\textit{Fermi}$-LAT blazar sources. Furthermore, we conduct temporal analysis of gamma-ray light curves and spectral analysis of blazar sources with the highest clustering of neutrino events.}
   {We find an excess of high-energy neutrino events from the directions of $\textit{Fermi}$ blazars beyond random coincidence, with a significance of 2.42~$\sigma$. Among the nine blazars which have a maximum number of neutrino events (four) arriving from within 3$^\circ$ from them, 4FGL~J1012.3+0629 exhibits the most significant variability around the neutrino detection time. Our study identifies 4FGL J1012.3+0629, 4FGL J2118.0+0019, 4FGL J2226.8+0051 and 4FGL J2227.9+0036 as plausible high-energy neutrino source candidates. The neutrino events IC110726A, IC170308A, IC190415A, IC141210A, IC150102A, IC230524A, IC140114A, IC110807A, and IC200523A likely originate from these sources. Additionally, we find that the spectral index tends to harden when these high-energy neutrino events are detected.}
   {}


   \maketitle
%
%-------------------------------------------------------------------

\section{Introduction}
The highest energy cosmic rays are believed to be accelerated in extragalactic astrophysical sources \citep{Kotera_2011}. When cosmic rays interact with surrounding matter and radiation in or near these sources, high-energy neutrinos are produced \citep{Aab_2017}. Unlike charged cosmic rays, neutrinos are neutral and interact very weakly, enabling them to travel through space without being deflected from their source \citep{Reines_1960}. This makes neutrinos a valuable probe for identifying the origin of cosmic rays. Among various extragalactic sources, blazars—active galactic nuclei (AGN) with relativistic jets pointed at small angles toward us \citep{Urry_1995}—are considered prime candidates for the production of high-energy cosmic rays \citep{Stecker_1991, Dimitrakoudis_2012}. Due to this orientation, blazars are also strong candidates for being extragalactic sources of high-energy neutrinos \citep{Mannheim_1993, Atoyan_2001, Dermer_2009, Murase_2012, Cerruti_2015, Rodrigues_2018, Murase_2023}.

High-energy neutrinos are produced through the decay of charged pions, which are formed during interactions between high-energy protons and either matter (\textit{p}\textit{p} interactions) \citep{Murase@2013} or photon fields (\textit{p}$\gamma$ interactions) \citep{Winter_2013}. In addition to charged pions, neutral pions (\(\pi^0\)) are also produced in these interactions. These neutral pions decay, primarily producing two $\gamma$-ray photons, making $\gamma$-rays valuable indicators of hadronic interactions in astrophysical environments. Hadronic models propose that neutrinos and $\gamma$-rays are generated through the decay of pions \citep{Mannheim_1993, Atoyan_2002, MUCKE_2003}, though $\gamma$-rays can also be generated via leptonic processes such as bremsstrahlung or inverse Compton scattering \citep{Garrappa_2024}. 

One of the major challenges in identifying the sources of high-energy neutrinos lies in the limited angular resolution of neutrino detectors. Additionally, the uneven sky coverage of most observing facilities further complicates the identification process \citep{Garrappa_2024}.

Recent multi-frequency campaigns on flaring blazars have shown significant correlations across different wavelengths. Observations suggest that TeV emissions are correlated with flaring events \citep{Aharonian_2009, Donnarumma_2009, Vercellone_2010}. The highest flux states of blazars are typically explained by flares \citep{Nalewajko_2013}. The advent of multimessenger astronomy has opened exciting prospects for investigating high-energy neutrinos. In 2013, diffuse neutrino flux was detected \citep{Aartsen_2013}, and since then, the origin of these astrophysical neutrinos remains one of the central unanswered questions.

Over the past decade, the IceCube Neutrino Observatory \footnote[1]{\url{https://icecube.wisc.edu/}} has detected numerous high-energy astrophysical neutrino events \citep{Aartsen_2017c}. On September 22, 2017, IceCube observed a track-like neutrino event, IceCube-170922A, with an energy exceeding 290 TeV \footnote[2]{\url{https://gcn.gsfc.nasa.gov/gcn3/21916.gcn3}}. This event was directionally coincident with the blazar TXS 0506+056 and showed a temporal coincidence with a $\gamma$-ray flare from the same source \citep{Aartsen_2018a}. TXS 0506+056 was identified as the first significant neutrino source \citep{AartsenM_2018, Finley_2019}. Following this discovery, a global multi-wavelength observational campaign was launched, involving telescopes such as MAGIC, H.E.S.S., VERITAS, $\textit{Fermi}$-LAT, $\textit{AGILE}$, and $\textit{Swift}$ XRT \citep{Aartsen_2018a}.

To identify plausible high-energy neutrino sources, IceCube has launched a target of opportunity program. This initiative allows for the rapid search for multiwavelength counterparts to high-energy neutrino track events  (\citealt{Aartsen_2017}). PKS 1424\textendash24 was identified as the first blazar candidate of the IceCube neutrino event, coinciding with its $\gamma$-ray flaring activity (\citealt{Kadler_2016}). Source candidates for the high-energy neutrino events IC170922A and IC141209A have been identified. The source of IC170922A has been identified as 4FGL J0509.4+0542 (TXS 0506+056) , while the source of IC141209A is 4FGL J1040.5+0617 (GB6 J1040+0617) (\citealt{Garrappa_2024}). The source 4FGL J1040.5+0617 was flaring in the \textit{Fermi}-LAT when IceCube-141209A was detected (\citealt{Garrappa_2019}). Source 4FGL J2226.8+0051  coincides with four distinct neutrino events (\citealt{Garrappa_2024}). \cite{Garrappa_2024} presented some neutrino events that coincided with the identified sources.

 During the neutrino alert, TXS 0506+056 emitted fewer $\gamma$-rays, indicating a low state, but the emitted $\gamma$-rays had higher energies, representing a harder $\gamma$-ray state. This low but hard $\gamma$-ray state aligns with the conditions needed for neutrino emission. During this low/hard state, there is hadronic emission, where interactions involving protons (or nuclei) play a key role in producing the observed $\gamma$-rays and potentially the associated neutrinos (\citealt{Padovani_2018}). Several blazars exhibited spectrum hardening during flares that coincided temporally with the arrival times of neutrinos (\citealt{Garrappa_2019, Giommi_2020_a_a, Liao_2022}). The source TXS 0506+056 exhibited a spectral index of $2.1\pm0.2$ during the detection of the neutrino event IC170922A, indicating a relatively hard spectrum (\citealt{AartsenM_2018}).

Several other potential associations between neutrinos and blazars have been identified at lower significance levels. Significant studies include those by \cite{Franckowiak_2020, Giommi_2020, Garrappa_2019, Krauss_2018, Kadler_2016}.
The spatial correlations between neutrino alerts and sources from various catalogs do not provide conclusive evidence of association, as the number of observed correlations aligns with what would be expected from random coincidences. As the number of real-time alerts increases, more $\gamma$-ray sources are being found to coincide with multiple neutrinos. However, many of these coincidences are due to the large uncertainties in the neutrinos' arrival directions. So, it is necessary to conduct detailed multiwavelength studies to identify potential source candidates (\citealt{Garrappa_2024}). Light curve follow-up observations of the nearest sources to each IceCube neutrino event are encouraged to study (\citealt{Abbasi_2023}). Light curves can be constructed with different time bins to study the $\gamma$-ray characteristics of the blazar sources.  \cite{Bhatta_Dhital_2020} and \cite{Rajput_2020} constructed weekly and monthly binned \textit{Fermi}-LAT light curves respectively to investigate the $\gamma$-ray variability in the blazar sources. \citep{Abdollahi_2020} constructed light curves using two-month and yearly time bins to study the variability of the sources. Several groups conducted follow-up observations and applied statistical methods. These methods were used to analyze and determine the nature of unassociated sources based on their $\gamma$-ray characteristics (\citealt{Hassan_2012, Doert_2014}). 

We search for further plausible high-energy neutrino sources using high-energy track like astrophysical IceCube neutrino events by observing the coincidence of these neutrinos with the $\gamma$-ray flaring in the light curve of the blazar sources. The complete catalog of IceCube neutrino event datasets of possible astrophysical origin is provided in electronic format on Harvard Dataverse\footnote[3]{\href{https://doi.org/10.7910/DVN/SCRUCD}{doi.org/10.7910/DVN/SCRUCD}}. This dataset catalog includes information on neutrino event ID, right ascension, declination, event energy, signal etc. This catalog contains muon track events produced by charge-current interactions of high-energy neutrinos (with energies above \textasciitilde 100 TeV) observed by the IceCube Neutrino Observatory, located deep beneath the geographic South Pole in Antarctica. At these energy levels, neutrino events are highly likely to originate from astrophysical sources. The best-fit direction is derived from an iterative maximum-likelihood scan of all potential event directions that is performed after the alert event is identified, and is published as an update for real-time alerts. Additional information on these alerts can be found in \cite{Abbasi_2023}. Track-like neutrino events at approximately TeV energies are reconstructed with a typical angular resolution of less than $1^\circ
$ (\citealt{Aartsen_2017}). There are 348 astrophysical IceCube neutrino events. Eight events that also triggered IceTop and are likely due to cosmic-ray showers are excluded from our analysis. So, we have taken 340 IceCube neutrino events in our analysis.
\cite{Abbasi_2023} utilized neutrino events of astrophysical origin from the period 2011-05-14 to 2020-12-22 and identified spatial correlations between neutrino alerts and \textit{Fermi}-LAT sources, while we used data from 2011-05-14 to 2023-10-14. There are about 2310 blazar sources that include both BL Lacertae objects (BL Lacs) and flat spectrum radio quasars (FSRQs). These sources can be found in the fits file `gll\_psc\_v35.fit' associated with  the fourth \textit{Fermi} LAT catalog (4FGL-DR4)\footnote[4]{\ \url{https://fermi.gsfc.nasa.gov/ssc/data/access/lat/14yr_catalog/}} of $\gamma$-ray Sources.
\cite{Franckowiak_2020} conducted a study that examines the correlation between $\gamma$-ray blazars and high-energy neutrinos. Our work focus on the identification of the origin of sources of high-energy astrophysical neutrino events and connection between flaring $\gamma$-ray sources with neutrino events. 

In this work, we search for the sources of high-energy astrophysical neutrino events. To achieve this, we look for the excess of neutrino events arriving from the blazar sources over the number of neutrino events which we would observe by random chance. For quantifying this, we use right ascension (RA) and the declination (Dec) distributions of the real high-energy neutrino events to generate background samples for comparison. This simple approach obviates the necessity of the otherwise detailed inclusion of exposures. We further investigate the blazar sources whose directions are nearby to the arrival directions of maximum of the high-energy neutrinos, by performing spectral and temporal analysis of the sources' $\gamma$-ray emission.

The paper is structured as follows: In Section 2, we present observations and  analysis techniques. In Section 3, we present our results from the search of excess of high-energy neutrino events from the direction of blazar sources, as well as those from the spectral and temporal analyses of the $\gamma$-ray light curves of the blazar sources the positions of which lie close to arrival directions of maximal number of high-energy neutrino events. This will help infer whether these blazar sources are plausible high-energy neutrino sources. 
In Section 4, discussions on the results and their possible implications are presented and lastly in Section 5, we summarize our key findings.

\section{Observations and Data Analysis}

In this section, we outline the methods employed for data analysis. First, we  generate random samples based on the distributions of right ascension and declination observed from real high-energy astrophysical neutrino event data to identify whether these neutrino events originate from the directions of blazar sources. Then, we use high-energy neutrino data to identify the coincident neutrino events with blazar sources at an angular separation of 3 degrees. Additionally, we perform  {\textit{Fermi}-LAT} data analysis for both temporal and spectral characterization of the sources. These analysis identify the sources of origin of high-energy astrophysical neutrino events, number of neutrino events and blazar sources that have directional coincidence (within 3 degrees) with the position of the blazar sources and neutrino events respectively, and the temporal coincidence of the maximum clustered neutrino events with the source's $\gamma$-ray flaring. 

\subsection{Neutrino observations and data analysis}

Our data sample consists of 340 IceCube astrophysical track like neutrino events. The LAT 14-year Source Catalog (4FGL-DR4)  contains 2,310 $\gamma$-ray blazar sources, as listed in the FITS file `gll\_psc\_v35'. To analyze the spatial coincidence between these blazar sources and IceCube neutrino events, we first select a blazar source from the LAT catalog and count the total number of neutrino events located within an angular separation of 3 degrees from the source position. We repeat this process for each of the 2,310 blazar sources in the catalog. The analysis gives the total number of blazar sources whose positions lie within 3 degrees of the arrival directions of neutrino events. In addition, it identifies the total number of astrophysical neutrino events that are spatially correlated with blazar positions among the 340 IceCube astrophysical neutrino events.\

From 340 astrophysical IceCube neutrino events, we construct histograms for both right ascension and declination. The right ascension histogram shows a nearly uniform distribution, while the declination histogram is non-uniform. We generate 340 events randomly following the RA and the declination distributions that we obtain from the real neutrino observations. To generate 340 events randomly, the non-uniform declination values are evaluated from the probability density function (PDF) derived from a kernel density estimation (KDE) of the actual declination values of the IceCube astrophysical neutrino events. The RA values are taken uniformly. We then use 340 randomly generated events to analyze their coincidence with blazar positions. Again, for each blazar source from the LAT catalog, we count the total number of coincident random events within a 3-degree angular separation from the source positions, repeating the process for all 2,310 blazars. This analysis determines the total number of coincident blazar sources. This process is iterated 1000 times for the computation of the significance.

We also generate 340 random events using uniform distributions for comparison.  For each blazar source, we search for uniformly generated random events within an angular separation of 3 degrees from the source position and calculate the total number of coinciding blazar sources with these random event positions.  This process is repeated 1000 times, and the total number of coincident blazar sources is calculated. 
 
 \subsection{\textit{Fermi}-LAT observations and data analysis}

The \textit{Fermi} Large Area Telescope (\textit{Fermi}-LAT) (\citealt{Atwood_2009}), is one of the two gamma-ray detectors onboard the \textit{Fermi Gamma-ray Space Telescope (Fermi)} mission. \textit{Fermi}-LAT is a pair-conversion $\gamma$ -ray telescope.
We use \textit{Fermi} LAT data from  54700 to 60370 MJD to construct long term $\gamma$-ray light curve. The \textit{Fermi}-LAT $\gamma$-ray data were obtained from the Fermi Science Support Center (FSSC) data server\footnote[5]{\url{https://fermi.gsfc.nasa.gov/cgi-bin/ssc/LAT/LATDataQuery.cgi}}. The analysis was conducted using the \textit{Fermi} Science Tools software package version v11r5p3\footnote[6]{\url{http://heasarc.gsfc.nasa.gov/ftools}}.
We utilized the \textit{Fermi}-LAT fourth source catalog (4FGL)
 along with the galactic diffuse emission model (gll\_iem\_v07.fits)\footnote[7] {\url{https://fermi.gsfc.nasa.gov/ssc/data/access/lat/BackgroundModels.html}} and the extra-galactic isotropic diffuse emission model (iso P8R3\_SOURCE \_V3\_v1.txt), to construct the model xml file. Following the application of selection criteria, including good time intervals, livetime, exposure map, and diffuse response of the instrument, each event was processed with the instrument response function (IRF) `P8R3\_SOURCE\_V3\_v1'. The model xml file contains numerous sources within the region of interest (ROI). The spectral parameters of these sources are optimized by likelihood analysis\footnote[8]{\url{http://fermi.gsfc.nasa.gov/ssc/data/analysis/documentation/Cicerone/Cicerone_Likelihood}}. Additionally, sources outside the ROI, generally fixed to their 4FGL catalog values, are included in the model xml file.
The process of acquiring spectral parameters and source significances involves conducting a maximum likelihood fit (\citealt{Mattox_1996}). This fitting procedure incorporates a model that encompasses known sources from the 4FGL catalog. In addition to the known sources from the 4FGL catalog, the model incorporates the Galactic diffuse emission and residual background. Pass 8 data\footnote[9]{\url{http://fermi.gsfc.nasa.gov/ssc/data/analysis/documentation/}} are utilized to employ the 'SOURCE' event class (evclass=128), as recommended by the \textit{Fermi} LAT collaboration for analyzing small regions of interest less than 25 deg.  (\citealt{bruel2018fermi}).
The test statistic (TS) is defined as 

\begin{equation}
    TS = -2 \left[ \ln(L_0) - \ln(L_1) \right] \,,
\end{equation}
Here, \( L_0 \) represents the maximum-likelihood value for the model under the null hypothesis (i.e., without an additional source), while \( L_1 \) represents the maximum-likelihood value for the model that includes an additional source at a specified location (\citealt{Mattox_1996}).

We utilized the `P8R3 SOURCE' event class, with `evclass' set to 128 and `evtype' set to each event class encompasses various event types, enabling us to select events based on different criteria.
The default setting for `evtype' is 3. This includes all event types that consist of both the front and back sections of the tracker. The data have been divided into spatial bins, and each pixel in the binning corresponds to an angular size of  $0.1^{^\circ}$. The light curves of the blazar sources are constructed over an energy range of 100 MeV to 300 GeV (\citealt{Chatterjee_2021}).  The methodology for this analysis is detailed in \cite{Prince_2018}. We extract source names and their positions from the 14-year LAT Source Catalog (4FGL-DR4). The FITS file `gll\_psc\_v35' in LAT 14-year Source catalog (4FGL-DR4) contains 2,310 $\gamma$-ray  blazar sources.\

We constructed light curves in the daily, weekly, and monthly time bins. In daily and weekly time binning, the flux values and corresponding test statistic (TS) values are very low, resulting to the flux values in the form of upper limits for many bins. So we could not clearly identify the flaring activity of the sources in the daily and weekly time binning light curves, and thus we studied each source in monthly time binning. The upper limits to the flux were determined where the test statistic (TS) was less than 4. In the light curves created in the monthly time bins, we look for the flaring activity and check if the high-energy astrophysical IceCube neutrino event detected at that particular time lies in the time frame of the flaring in the source. Connection between the high variability in a source's emission and the time of neutrino detection is useful to infer whether the IceCube neutrino event is possibly associated with the source.

To compare the variability between selected sources, we find the fractional variability amplitude (\( F_{\text{var}} \)) which is a commonly used statistical measure and quantifies the intrinsic variability of a source relative to its mean flux while accounting for measurement uncertainties.

The fractional variability amplitude is defined as

\begin{equation}
    F_{\text{var}} = \sqrt{\frac{S^2 - \langle \bar{\sigma}^2_{\text{err}} \rangle}{\langle \bar{x} \rangle^2}}
    \label{eq:F_var}
\end{equation}
where \( S^2 \) is the sample variance of the measured flux values, \( \langle \bar{\sigma}^2_{\text{err}} \rangle \) is the  mean square error, and  \( \langle \bar{x} \rangle \) is the mean value of the sample.
The error in fractional variability, \( \text{err}(F_{\text{var}}) \), is given by

\begin{equation}
    \text{err}(F_{\text{var}}) = \sqrt{\left( \sqrt{\frac{1}{2N}} \frac{\langle \bar{\sigma}_{\rm err} \rangle^2}{\langle \bar{x} \rangle^2 F_{\text{var}}} \right)^2 
    + \left( \sqrt{\frac{\langle \bar{\sigma}_{\rm err}^2 \rangle}{N}} \frac{1}{\langle \bar{x} \rangle} \right)^2 }.
    \label{eqn:deltaNXS}
\end{equation}

\( N \) is the total number of observations.
A detailed discussion on the fractional variability can be found in \citep{Vaughan_2003, Rani_2016}.

We analyze the hardness or softness of the photon and spectral indices, which provide valuable information into the potential connection between high-energy neutrinos and gamma-ray activity. The photon index is associated with the Powerlaw model, where as the spectral index is related to the Logparabola model.
The Powerlaw model (\citealt{Fan_2002, Fan_2013, Abdo_2010b}) can be written as

\begin{equation}
    \dfrac{dN}{dE}=NE^{-\Gamma}
\end{equation}
where N and $\Gamma$ are normalization constant and photon index, respectively.\
The Logparabola model (\citealt{Abdalla_2017, Dermer_2015}) can be expressed as

\begin{equation}
    \frac{dN}{dE} = N_o (\frac{E}{E_o})^{-(\alpha + \beta \log\left(\frac{E}{E_o}\right))}
\end{equation}
Where $N_{0}$, ${E_0}$, $\alpha$, and $\beta$  are the normalization constant, peak energy, spectral index and curvature parameter respectively.

\section{Results}

\subsection{Excess in the number of high-energy neutrino events from the direction of blazar sources}
\begin{figure}[htbp]
  \centering
  
  \includegraphics[width=\hsize]{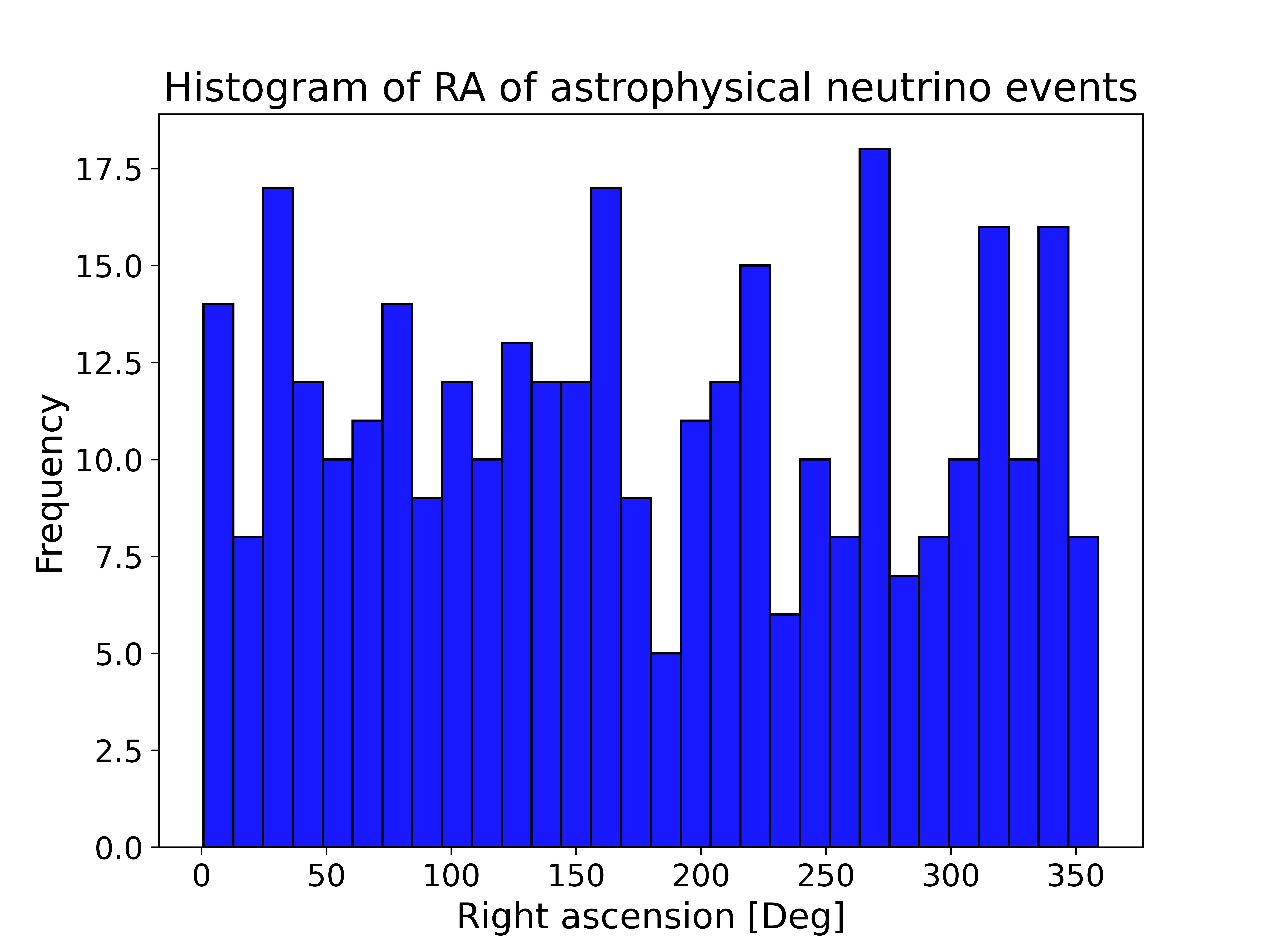}
  \caption{Histogram of right ascension (RA) of high-energy IceCube astrophysical neutrino events. RA histogram shows that the neutrino events are nearly uniformly distributed.} \label{fig1}
\end{figure}

\begin{figure}[htbp]
  \centering
  
  \includegraphics[width=\hsize]{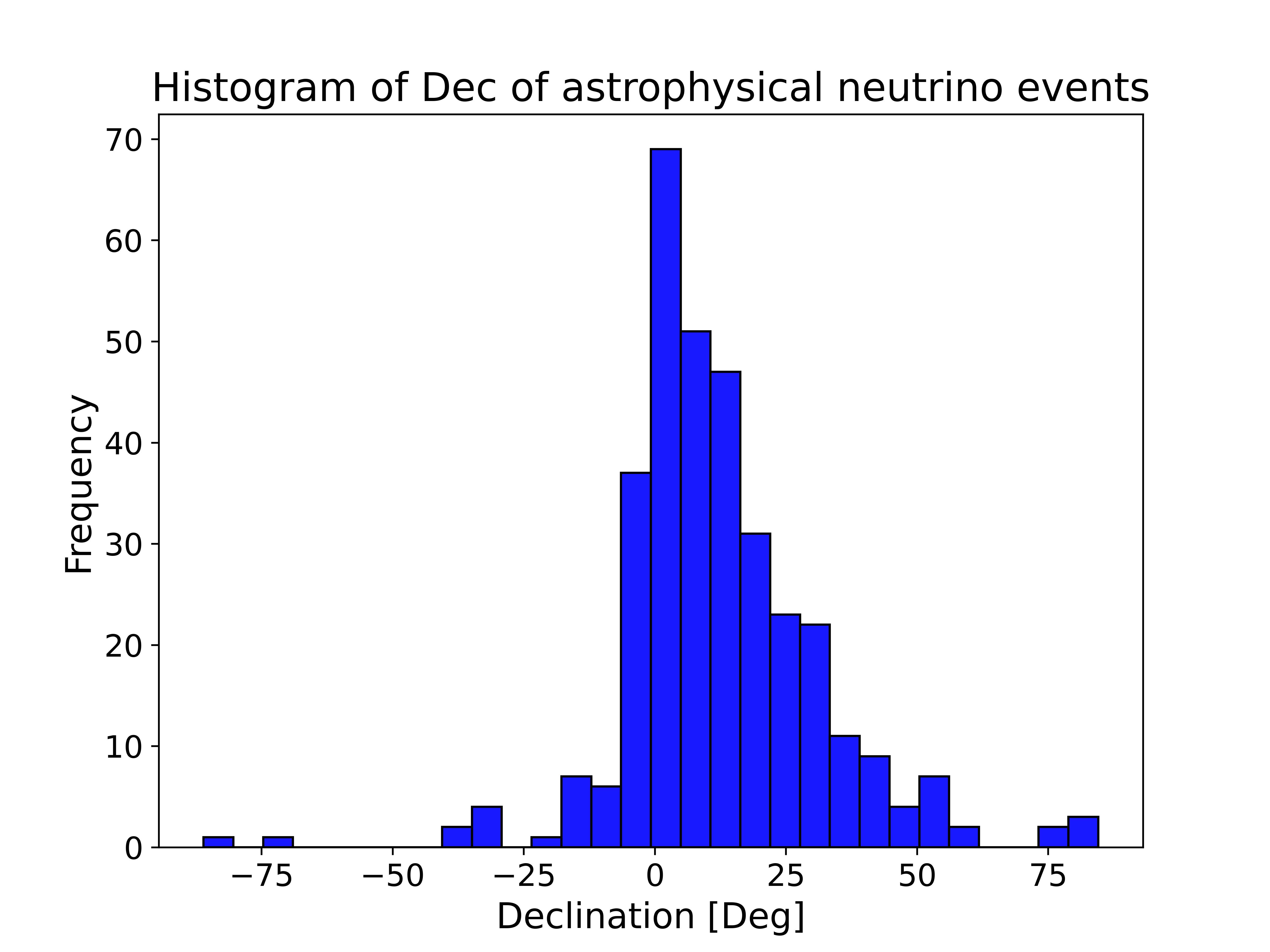}
\caption{Histogram of declination (Dec) of the high-energy IceCube astrophysical neutrino events. Declination histogram shows that the neutrino events are non-uniformly distributed.}
\label{fig2}
\end{figure}

The histogram of right ascensions corresponding to the arrival directions of the high-energy neutrino events reveals a uniform distribution as shown in Figure \ref{fig1},  as expected, while the histogram of the corresponding declinations (see Figure \ref{fig2}) follow a non-uniform distribution which is due to the fact that the events from the southern hemisphere are heavily suppressed during the event filtering, and  upgoing neutrinos are more effectively detected as the Earth filters out atmospheric muons.\ 

The similarity in the overall shape of the two histograms (see Figure \ref{fig2} and Figure  \ref{figA1A2} (right panel)) validates the method used to generate the random non-uniform declination values. This ensures that the generated data accurately reflects the declination trends observed in the IceCube neutrino data. The observed neutrino events in declination distribution are consistent with a modeled random non-uniform distribution of declinations, which was derived based on the declination distribution of the real astrophysical neutrino data.
The declination distribution clearly reflects the effect of reduced sensitivity of the IceCube to events originating from the southern hemisphere.\
  
\begin{figure}[htbp]
  \centering
  
   \includegraphics[width=\hsize]{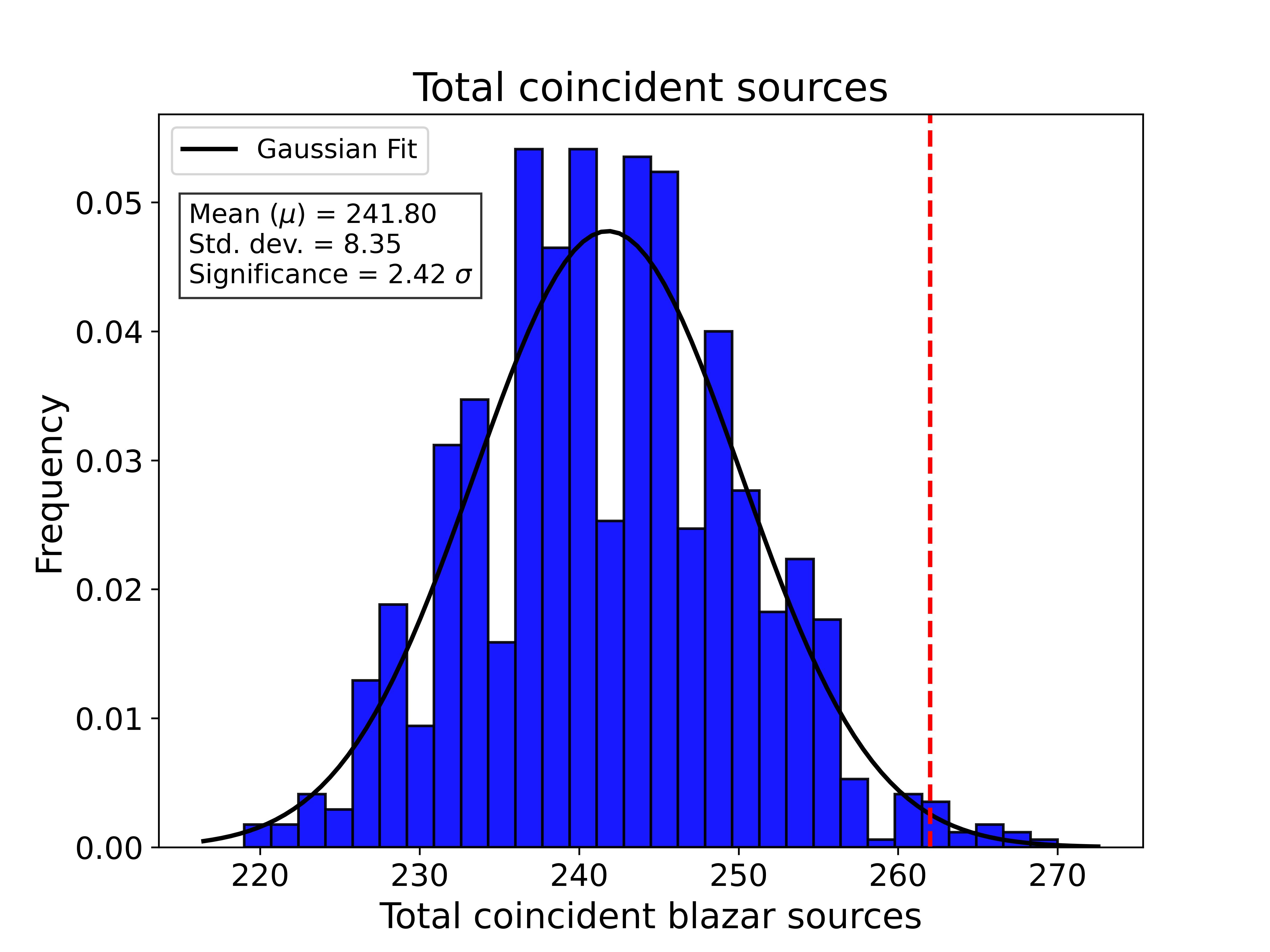}

\caption{Histogram of total counts of high-energy neutrino events which have their simulated arrival directions contained within 3 degrees of \textit{Fermi} LAT blazar sources. Sampling of the arrival directions was performed using the distributions of RA and declination of real neutrino events. Also, a red dashed vertical line corresponding to the total counts of real neutrino events is shown.}\label{fig3}
\end{figure}

To find out if there is any excess in the number of high-energy neutrino events arriving from the directions of blazars, we simulate 1000 samples of events, each sample comprising 340 events which is the number of real high-energy neutrino events in our analysis, with their right ascensions and declinations randomly sampled from the distributions of right ascension and declination of the real events. For a comparison, we also used a uniform distribution for declinations for sampling. Using uniform distributions for both the RA and Dec, an average of 237 events were found to be arriving from within 3 degrees of blazar sources included in the LAT 14-year Source Catalog (see Figure \ref{figA1A2}(left panel)). Also, using the distributions of RA and Dec from the data for sampling, we obtained an average of 242 such events as shown in Figure \ref{fig3}. This number is what one would expect if the high-energy neutrino events were arriving to the detector if the origins of these neutrinos were not related with the considered blazar sources. However, from the data used in our analysis there are 262 events (shown in Figure \ref{fig3} by dashed red line) arriving from within 3 degrees of the \textit{Fermi} LAT blazar sources, which corresponds to an excess with a significance of 2.42 $\sigma$.\
 
\subsection{Identification of neutrino sources from light curve follow up observations}

 Among 340 neutrino events, we found that 262 neutrino events have arrival directions lying within 3 degrees from the directions of 460 \textit{Fermi} LAT blazar sources. Among  460 sources, 362 blazar sources contain a single neutrino event within an angular separation of 3 degrees from the source position while 73 blazar sources contain two neutrino events. Also, there are 16 sources containing three neutrino events and 9 sources containing 4 neutrino events each. Among these nine blazar sources, three sources— 4FGL J2223.3+0102 (RA: 335.85$^{\circ}$, DEC: 1.05$^{\circ}$), 4FGL J2226.8+0051 (RA: 336.71$^{\circ}$, DEC: 0.86$^{\circ}$), and 4FGL J2227.9+0036 (RA: 336.98$^{\circ}$, DEC: 0.62$^{\circ}$) share same neutrino events, and two sources 4FGL J1016.0+0512 (RA: 154.01$^{\circ}$, DEC: 5.21$^{\circ}$) and  4FGL J1018.4+0528 (RA: 154.62$^{\circ}$, DEC: 5.47$^{\circ}$) share same neutrino events. It is uncertain whether these neutrinos originate genuinely from these sources. For this we investigate the $\gamma$-ray/neutrino connection using the $\gamma$-ray light curves of the blazars to determine if these are plausible candidate sources of the neutrino events too. We perform both  timing and spectral analyses of the $\gamma$-ray emission of these blazar sources.
 In Table \ref{table1}, we present the list of nine blazar sources with the highest number of neutrino events detected around them within an angular separation of 3 degrees. The table also provides the spatial positions of the neutrino events and blazar sources, along with the angular separations between them. Similarly, among the nine blazar sources analyzed, we identified four sources as plausible candidates for high-energy astrophysical neutrinos. These potential source candidates, along with their neutrino events that lie in the significant flux state in the monthly time binning of the long term $\gamma$-ray light curves of the sources, are presented in Table \ref{table2}. The $\gamma$-ray light curves of these plausible sources for high-energy astrophysical neutrino events are presented in Figures \ref{fig5}, \ref{fig6}, \ref{fig7}, and \ref{fig8}.

\begin{table}[htbp]
   \caption[]{Nine blazar sources each containing four neutrino events, showing source names, their positions, neutrino events, their positions, and angular separations. Each values of RA, Dec and separation (Sep) are expressed in degrees.}\label{table1}
   
   \begin{tabular}{p{0.37\linewidth} l}
   
      \hline
      \noalign{\smallskip}
      \textbf{Source (RA$^{\circ}$, Dec$^{\circ}$)} & \textbf{Event (RA$^{\circ}$, Dec$^{\circ}$, Sep $^{\circ}$)} \\
      \noalign{\smallskip}
      \hline
      \noalign{\smallskip}
      \text{4FGL J0506.9+0323} & \text{IC161117A (78.66, 1.60, 2.63)} \\
      \text{(76.73, 3.39)} & \text{IC170922A (77.43, 5.79, 2.50)} \\
                             & \text{IC181008A (77.08, 1.23, 2.19)} \\
                             & \text{IC220918A (75.15, 3.58, 1.59)} \\
      \noalign{\smallskip}
      \hline
      \text{4FGL J1012.3+0629} & \text{IC110726A (151.08, 6.99, 2.05)} \\
      \text{(153.08, 6.50)} & \text{IC150118A (152.53, 4.33, 2.23)} \\
                             & \text{IC170308A (155.35, 5.53, 2.45)} \\
                             & \text{IC190415A (154.86, 5.27, 2.15)} \\
      \noalign{\smallskip}
      \hline
      \text{4FGL J1016.0+0512} & \text{IC130627B (155.35, 3.73, 1.99)} \\
      \text{(154.01, 5.21)} & \text{IC150118A (152.53, 4.33, 1.72)} \\
                             & \text{IC170308A (155.35, 5.53, 1.37)} \\
                             & \text{IC190415A (154.86, 5.27, 0.85)} \\
      \noalign{\smallskip}
      \hline
      \text{4FGL J1018.4+0528} & \text{IC130627B (155.35, 3.73, 1.89)} \\
      \text{(154.62, 5.47)} & \text{IC150118A (152.53, 4.33, 2.37)} \\
                             & \text{IC170308A (155.35, 5.53, 0.73)} \\
                             & \text{IC190415A (154.86, 5.27, 0.31)} \\
      \noalign{\smallskip}
      \hline
      \text{4FGL J2118.0+0019} & \text{IC130509A (317.50, 2.09, 2.67)} \\
      \text{(319.50, 0.33)} & \text{IC141210A (318.12, 1.57, 1.86)} \\
                             & \text{IC150102A (318.74, 2.91, 2.69)} \\
                             & \text{IC230524A (318.43, 2.84, 2.73)} \\
      \noalign{\smallskip}
      \hline
      \text{4FGL J2223.3+0102} & \text{IC110807A (336.80, 1.53, 1.07)} \\
      \text{(335.85, 1.05)} & \text{IC140114A (337.59, 0.71, 1.77)} \\
                             & \text{IC200523A (338.64, 1.75, 2.88)} \\
                             & \text{IC221224A (335.74, 1.42, 0.39)} \\
      \noalign{\smallskip}
      \hline
      \text{4FGL J2226.8+0051} & \text{IC110807A (336.80, 1.53, 0.67)} \\
      \text{(336.71, 0.86)} & \text{IC140114A (337.59, 0.71, 0.89)} \\
                             & \text{IC200523A (338.64, 1.75, 2.12)} \\
                             & \text{IC221224A (335.74, 1.42, 1.12)} \\
      \noalign{\smallskip}
      \hline
      \text{4FGL J2227.9+0036} & \text{IC110807A (336.80, 1.53, 0.93)} \\
      \text{(336.98, 0.62)} & \text{IC140114A (337.59, 0.71, 0.61)} \\
                             & \text{IC200523A (338.64, 1.75, 2.01)} \\
                             & \text{IC221224A (335.74, 1.42, 1.48)} \\
      \noalign{\smallskip}
      \hline
      \text{4FGL J2252.6+1245} & \text{IC120523B (343.78, 15.48, 2.79)} \\
      \text{(343.17, 12.75)} & \text{IC190619A (343.52, 10.28, 2.50)} \\
                             & \text{IC210629A (340.75, 12.94, 2.36)} \\
                             & \text{IC230201A (345.41, 12.10, 2.29)} \\
      \noalign{\smallskip}
      \hline
   \end{tabular}
    \begin{flushleft}
      \footnotesize \textbf{Note:} Sep$^{\circ}$ represents angular separation between the blazar source and the neutrino event in degrees.
   \end{flushleft}
\end{table}

\begin{figure}[htbp]
 \centering
   \includegraphics[width=\hsize]{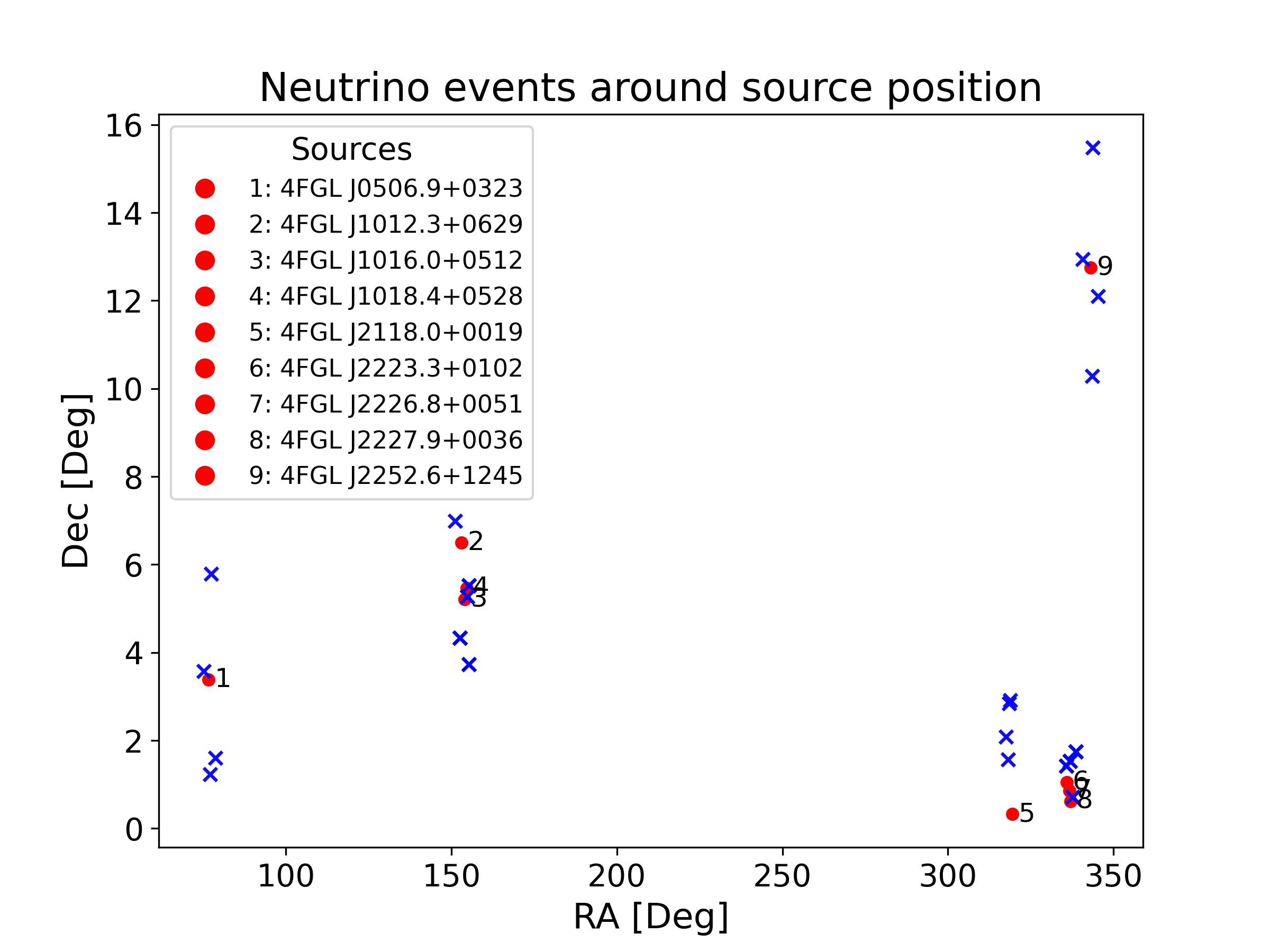}

  \caption{RA and Dec plot of nine blazar sources, each containing four neutrino events within an angular separation of 3 degrees.}
\end{figure}

\begin{figure}[htbp]
  \centering
  
  \includegraphics[width=\hsize]{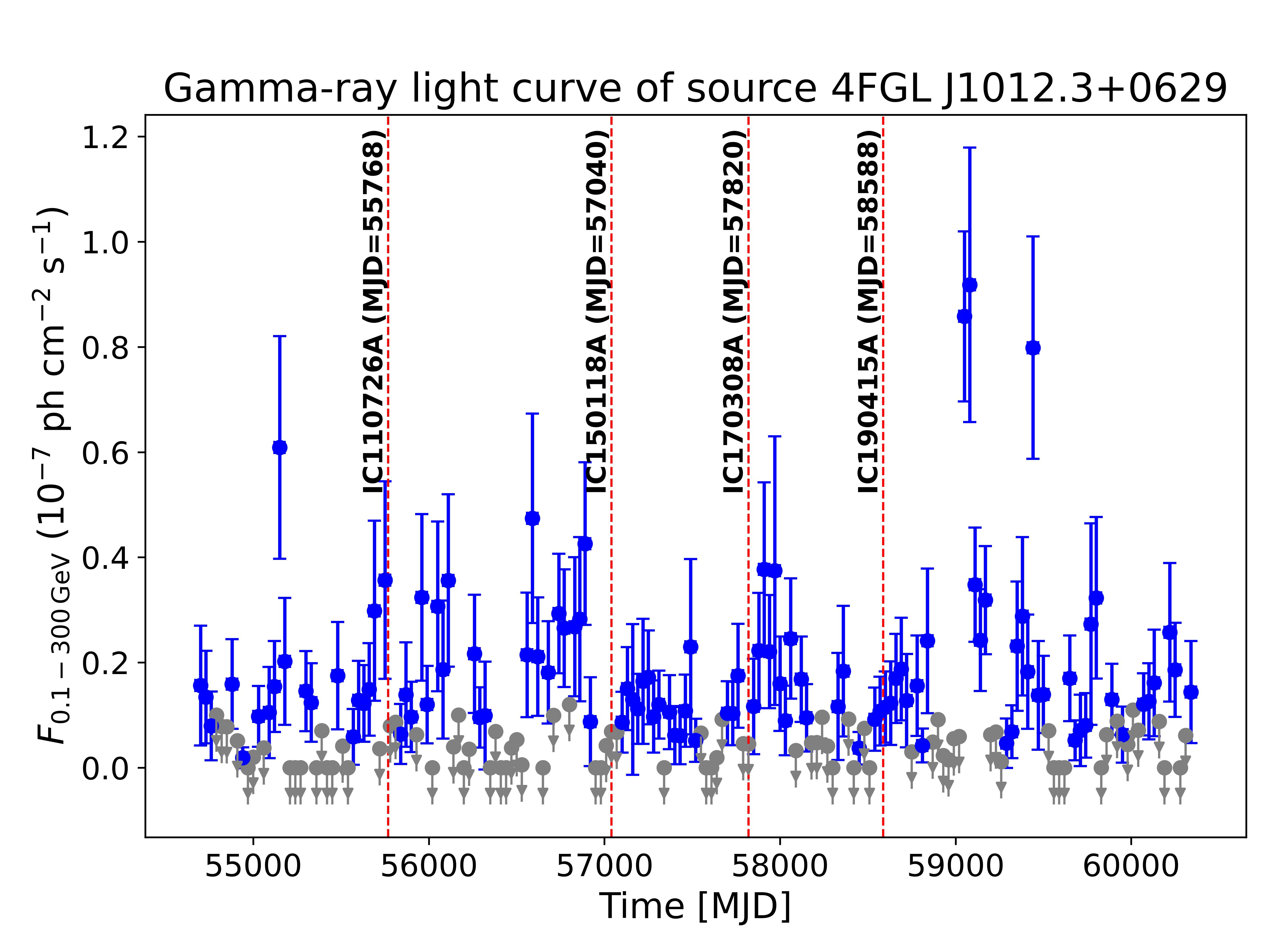}
  \caption{$\gamma$-ray light curve of the source 4FGL J1012.3+0629 from  54700 to 60370 MJD in 30-days time bins from 100 MeV to 300 GeV. Red dashed lines represents the detection time of IceCube neutrino events.} \label{fig5}

\end{figure}

\begin{figure}[htbp]
  \centering
  
  \includegraphics[width=\hsize]{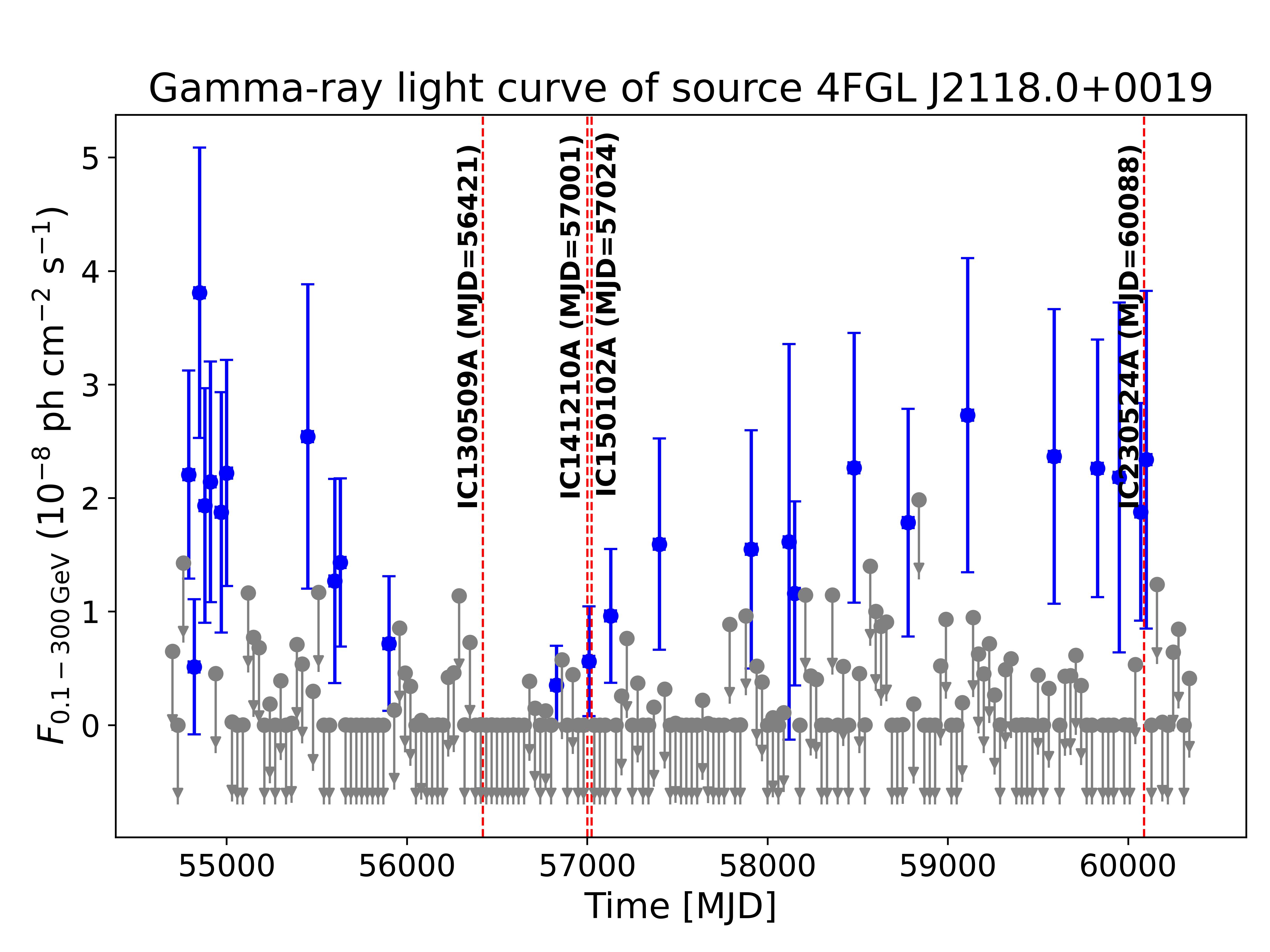}
  \caption{$\gamma$-ray light curve of the source 4FGL J2118.0+0019 from  54700 to 60370 MJD  in 30-days time bins.} \label{fig6}
\end{figure}

\begin{figure}[htbp]
  \centering
  
  \includegraphics[width=\hsize]{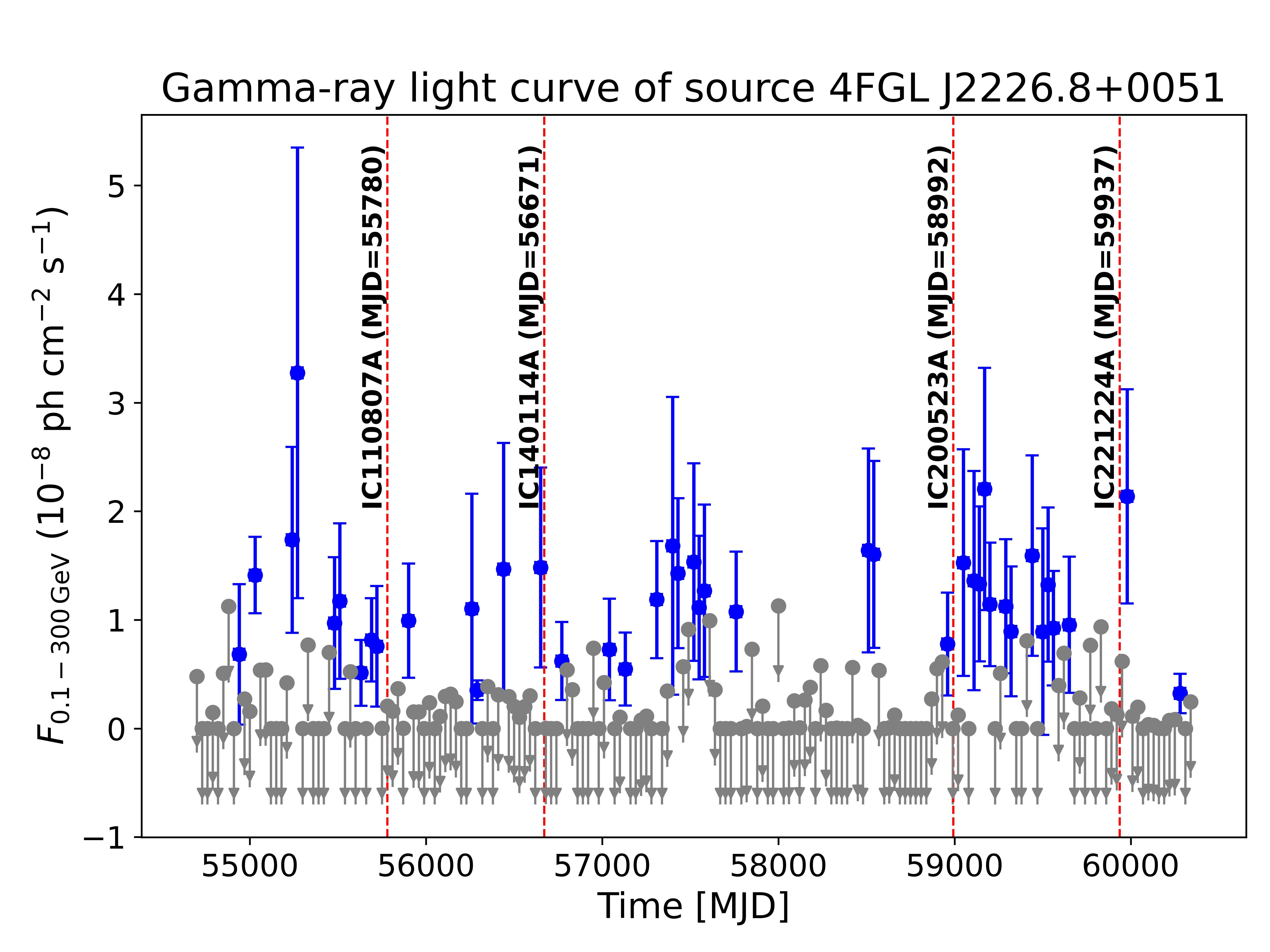}
  \caption{$\gamma$-ray light curve of the source 4FGL J2226.8+0051  from  54700 to 60370 MJD in 30-days time bins.} \label{fig7}
\end{figure}

\begin{figure}[h!]
  \centering
  
  \includegraphics[width=\hsize]{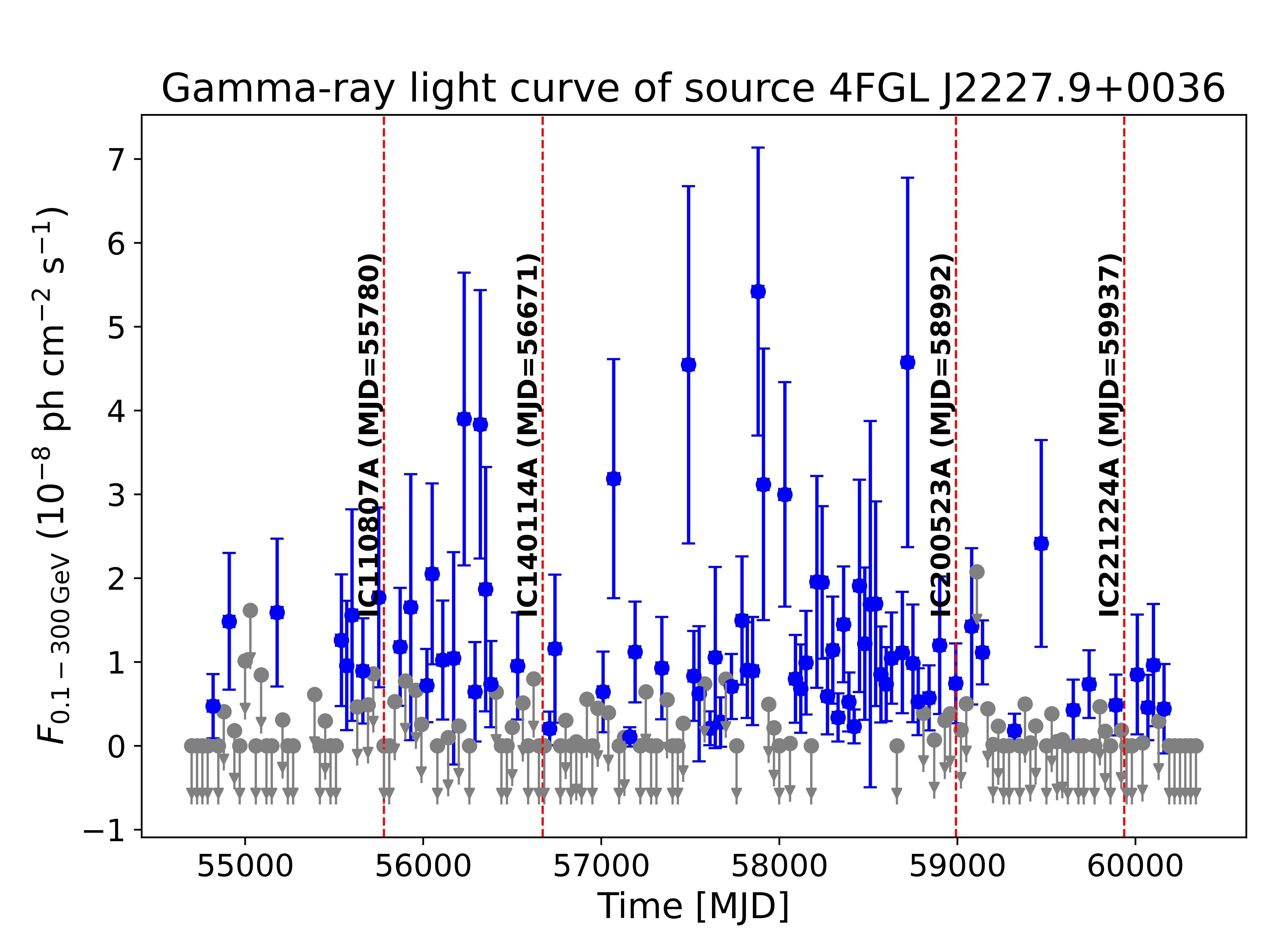}
  \caption{$\gamma$-ray light curve of the source 4FGL J2227.9+0036 from  54700 to 60370 MJD  in 30-days time bins.} \label{fig8}
\end{figure}

\begin{table}[htbp]
    \centering
    \caption{Coincident neutrino events with significant flux state in the light curve of the source.}\label{table2}
    \begin{tabular}{cc}
    \hline
    \textbf{Source} & \textbf{Neutrino events} \\
     \hline
   4FGL J1012.3+0629 & IC110726A, IC170308A, IC190415A  \\
    4FGL J2118.0+0019 &IC141210A, IC150102A, IC230524A  \\
    4FGL J2226.8+0051 &IC140114A  \\
    4FGL J2227.9+0036 &IC110807A, IC200523A  \\
   
    \hline
    \end{tabular}
\end{table}

\section{Discussion}
The number of neutrino events observed at an angular separation of less than 3$^{\circ}$  from the positions of the \textit{Fermi} LAT blazar sources is 262, which is higher than the average value that would be expected based purely on random fluctuations. To quantify this excess, statistical significance is used, which compares the observed number of events to what would be predicted if there were no actual association between the astrophysical neutrino events and the blazar sources. The significance of 2.42 $\sigma$ indicates that the excess is unlikely to be due to random fluctuation but could be a real signal. However, because 2.42 $\sigma$ is below the usual threshold for discovery, this result is considered significant but not definitive. The observed excess could be due to a real physical connection between the neutrinos and the blazar sources, possibly due to hadronic interactions occurring within these astrophysical objects. These interactions can produce neutrinos in addition to $\gamma$ rays, supporting models of blazars as potential cosmic accelerators, but further analysis is needed to confirm this.

\subsection{Timing Analysis}
In our timing analysis, we studied whether the sources exhibited flaring activity or a higher-than-usual emission around the neutrino detection time for which we constructed long term $\gamma$-ray light curves in monthly time binning from 54700 to 60370 MJD of these nine blazar sources that contain four neutrino events around it. \

The neutrino event IC110726A falls within the monthly time regime of the high flux state of the source 4FGL J1012.3+0629. There is clear flaring activity in the source around the detection time of neutrino events IC110726A, IC170308A and IC190415A. This indicates that the neutrino events IC110726A, IC170308A, and  IC190415A likely originate from the source 4FGL J1012.3+0629  (see Figure \ref{fig5}).\

For the source 4FGL J2118.0+0019, the neutrino event IC130509A, detected on 56421 MJD and arriving at an angular separation of 3 degrees from the position of the source, lies within a monthly time regime that does not exhibit significant (TS $\geq 4$) photon flux. The neutrino events IC141210A and IC150102A are observed during photon flux state that are elevated compared to the nearby flux values, which are represented as upper limits because of their low significance. Both neutrino events IC141210A
and IC150102A fall in the same state of significant photon
flux. This significant photon flux state may suggest the presence of additional energy during these periods relative to the surrounding time bins. The alignment of these neutrino events with a significant photon flux state is notable, particularly given the prevalence of upper limits before and after these periods, indicating a relatively significant flux state at the time of these events. However, this observation highlights the need for further exploration of the underlying processes that could contribute to this elevated photon flux level. The neutrino event IC230524A, detected on 60088 MJD, aligns with a period of elevated photon flux in the $\gamma$-ray light curve of the source 4FGL J2118.0+0019. The neutrino
event IC230524A aligns with two instances of significant
photon flux observed in consecutive regular time bins, indicating a correlation between the arrival time of the neutrino
event and $\gamma$-ray activity. This elevated flux is notable when compared to adjacent time bins, which are represented by upper limits. The elevated photon flux connected with the neutrino event IC230524A detection time may be indicative of additional energy processes that occurred during this period.\

For the source 4FGL J2226.8+0051, the neutrino events IC110807A, IC200523A, and IC221224A are observed during periods of low photon flux, represented by upper limits because of their low significance. In contrast, the neutrino event IC140114A aligns with a time bin that has significant photon flux. This suggests a potential correspondence between the detection time of IC140114A and an elevated photon flux in the monthly time binning. \

For the source 4FGL J2227.9+0036, two neutrino events IC110807A, IC200523A lie in the time bins that have significant photon flux, while the neutrino events IC140114A and IC221224A lie in the time bins of those photon flux which are represented by upper limits.\

 The source 4FGL J1016.0+0512  are primarily driven by its active state during the initial phase, which contributes significantly to its overall high value of fractional variability among the selected sources. We observe the same conditions for the source 4FGL J1018.4+0528. The sources with their corresponding fractional variability values are presented in Table \ref{table3}. \

 \begin{table}[htbp]
   \caption[]{Source and their corresponding fractional variability values}\label{table3}
   \label{tab:your_table_label} 
   
   \begin{tabular}{lc}
      \hline
      \noalign{\smallskip}
      \textbf{Source}  & \textbf{Fractional variability ($F_{\text{var}}$)} \\
      \noalign{\smallskip}
      \hline
      \noalign{\smallskip}
      4FGL J0506.9+0323  & $0.91 \pm 0.41$ \\
      \noalign{\smallskip}
      4FGL J1012.3+0629  & $0.83 \pm 0.20$ \\
      \noalign{\smallskip}
      4FGL J1016.0+0512  & $1.84 \pm 0.06$ \\
      \noalign{\smallskip}
      4FGL J1018.4+0528  & $1.10 \pm 0.11$ \\
      \noalign{\smallskip}
      4FGL J2118.0+0019  & $0.11 \pm 0.50$ \\
      \noalign{\smallskip}
      4FGL J2223.3+0102  & $0.00 \pm 0.00$ \\
      \noalign{\smallskip}
      4FGL J2226.8+0051  & $ 0.43 \pm 0.13$ \\
      \noalign{\smallskip}
      4FGL J2227.9+0036  & $0.88 \pm 0.08$ \\
      \noalign{\smallskip}
      4FGL J2252.6+1245 & $0.78 \pm 0.17$ \\
      \noalign{\smallskip}
      \hline
   \end{tabular}
   
   \begin{flushleft}
     
   \end{flushleft}
\end{table}

For the sources 4FGL J0506.9+0323, 4FGL J1016.0+0512, 4FGL J1018.4+0528, 4FGL J2223.3+0102, and 4FGL J2252.6+1245, no significant photon flux in the monthly time bins aligns temporally with the detection of the neutrino event. 
Therefore, we cannot conclude that any of the neutrino events, despite their spatial coincidence within a 3$^\circ$ angular separation from these \textit{Fermi}-LAT blazars, originate from these sources. The light curves of these sources, which are unlikely to be plausible origins of high-energy neutrino events, are presented in Appendix \hyperref[appendix:C]{C}.\

We identified four \textit{Fermi}-LAT blazar sources—4FGL J1012.3+0629, 4FGL J2118.0+0019, 4FGL J2226.8+0051, and 4FGL J2227.9+0036 as plausible candidates for high-energy neutrino sources. The association between these blazar sources and neutrino events, shown in Table \ref{table2}, is inferred based on the temporal alignment of neutrino events with significant photon flux (TS $\geq 4$) obtained through monthly time-bin integration.\

The peak flux of the blazars we identified as plausible source candidates is comparable to the values reported for other neutrino source candidate blazars (\citealt{Liao_2022}). This similarity provides strong evidence supporting the plausibility that the sources we identified are additional members of the high-energy neutrino sources.\

For a direct comparison of the variability across the selected sources that have the maximum number of neutrino events, we have presented photon flux and variability index (\citealt{Abdollahi_2020, Itoh_2016}). These values are obtained from the 4FGL-DR4 catalog and represent the integrated measurements during the observation period from 4 August 2008 to 2 August 2022, spanning 14 years (see Appendix \hyperref[appendix:D]{D}). These integrated values serve as a reference to characterize the temporal variability of the selected sources. The data presented in Appendix \hyperref[appendix:D]{D} enables a comparative analysis of variability across the sources, offering a clear basis for comparison.\

Significant photon flux measurement coinciding with a neutrino event in the monthly time bin is not sufficient evidence to definitively conclude that particular photon flux is responsible for neutrino production. Temporal correlation alone does not prove causation. Although  significant photon flux and neutrino event occur in the same time bin, this could be coincidental. To establish a connection between significant photon flux and neutrino production, we perform an analysis of the source's spectral properties, focusing on its spectral behavior like hardness and softness, which reveal the energy distribution within the spectrum. Thus, studying the spectral behavior may help identify whether a source is a potential emitter of high-energy neutrinos. We analyze whether the photon index and spectral index exhibit hardening or softening, which provides an understanding of the efficiency of particle acceleration within the source. 

\subsection{Spectral Analysis}
The spectral parameters of all nine sources across different time episodes, analyzed using monthly time binning, are presented in Appendix \hyperref[appendix:B]{B}. We analyze the spectral index and photon flux only for plausible sources identified through temporal analysis, focusing on the time episodes where neutrino events align with the high flux (flaring) states of the sources.\ 

For the source 4FGL J1012.3+0629, the spectral index $-2.00 \pm 0.06$ during the period 58300 to 59200 MJD is the hardest compared to other time periods. The spectral index $-2.00 \pm 0.06$ is hardest when the neutrino event IC190415A is detected. The observed photon flux at this value of the spectral index is $(17.04\pm2.05)$ $\times 10^{-9}$ cm$^{-2}$ s$^{-1}$ which is highest among other observations calculated for the source 4FGL J1012.3+0629. This indicates that the blazar is in an active state and is consistent with flaring or outburst activity in the jet. The particle population is mostly made up of newly accelerated particles that have not lost much energy, which keeps the spectrum hard and results in higher flux. Similarly a harder $-2.07 \pm 0.07$ spectral index with a photon flux value $(13.08 \pm 2.00) \times 10^{-9}$ cm$^{-2}$ s$^{-1}$ is found when the neutrino event IC170308A is detected. In addition, a relatively harder spectral index of $-2.11 \pm 0.09$ is recorded in the time episode when the neutrino event IC110726A is detected. This hardening of the spectral index may indicate changes in particle acceleration or energy distribution, possibly as a result of varying physical conditions within the source. \
 
For the source 4FGL J2118.0+0019, the hardest spectral index $-2.01\pm 0.45$ with the least photon flux value $(0.98\pm1.26)$ $\times 10^{-9}$ cm$^{-2}$ s$^{-1}$  is observed during the time episode 56500–57400 MJD where we find two neutrino events IC141210A and IC150102A occurring within a monthly time frame characterized by significant photon flux. This suggests that fewer high-energy photons are emitted, potentially indicating efficient particle acceleration. These two neutrino events occurring on a shorter time scale and aligned with the same significant photon flux are likely the result of efficient particle acceleration.\ 

Similarly, the spectral index $-2.26 \pm 0.27$ is relatively harder when the high-energy neutrino event IC230524A is detected. This hardening of the spectral index indicates that $\gamma$-ray photons are being produced by the acceleration of charged particles to higher energies.\ 

For the source 4FGL J2226.8+0051, the spectral index ($2.34 \pm 0.54$) is found to be relatively harder with a lower photon flux value of $(1.75\pm1.81)$ $\times 10^{-9}$ cm$^{-2}$ s$^{-1}$ when the IceCube neutrino event IC140114A, detected on 56671 MJD, falls within a time zone having significant photon flux state. The source is likely in a low-activity phase, producing a relatively small amount of high-energy photons but with a harder spectral slope. \

For the source 4FGL J2227.9+0036, a relatively harder spectral index of ($-1.99 \pm 0.09$) is measured over the time range 55600 to 56500 MJD, within which the IceCube neutrino event IC110807A, detected specifically on MJD 55780, aligns with a time bin that exhibits significant photon flux. Also, a relatively harder spectral index ($-1.93 \pm 0.08$) is measured in the time range 58300 to 59200 when the neutrino event IC200523A is detected. This harder spectral index suggests that photons in this period are likely produced by the acceleration of charged particles. The relatively harder spectral index suggests an excess of photon flux in higher energy regime. This low/hard $\gamma$-ray state of the source during the neutrino detection time can be due to the fact that these neutrino events likely originate from these sources.\

The spectral analysis of the blazars shown in the tables of Appendix \hyperref[appendix:B]{B} reveals indications of both 'harder-when-brighter' and 'softer-when-brighter' trends. The 'harder-when-brighter' trend refers to a phenomenon observed in blazars where the emitted spectrum shifts toward higher energies (becomes harder) as the brightness increases. Conversely, the 'softer-when-brighter' trend occurs when the spectrum shifts toward lower energies (becomes softer) with increasing brightness. These trends have been widely observed across various frequency bands, including $\gamma$-ray, during fluctuations in blazar flux over time (see, e.g., \citealt{Chatterjee_2023, Noel_2022, Bhatta_2018}). Fluctuations in the blazar flux over time are shown in the tables of Appendix \hyperref[appendix:B]{B}. During the 'harder-when-brighter' trend, the photon index becomes flatter (harder spectrum) as the flux increases. It indicates efficient acceleration of high-energy protons in the jet. During these phases, protons can interact with ambient photons or gas, producing pions and subsequently high-energy neutrinos. Flares with harder spectra may signal enhanced conditions for neutrino production, as the same processes that accelerate protons to high energies also create conditions for \textit{p}\textit{$\gamma$} or \textit{p}\textit{p} interactions.\

The spectra of various sources across different time intervals fit the powerlaw (PL)  and log-parabola (LP)   models and are shown in Appendix \hyperref[appendix:B]{B}. These models describe non-thermal radiation from relativistic electrons within the magnetic field of a blazar jet. In the synchrotron emission framework, a PL spectrum represents coherent synchrotron radiation arising from a population of particles following a simple powerlaw energy distribution. But, log-parabolic models account for more complex scenarios, where the energy spectrum of relativistic electrons is shaped by statistical acceleration mechanisms. These mechanisms introduce an energy-dependent probability of acceleration, leading to the curvature observed in the log-parabolic spectrum (\citealt{Massaro_2006}).\

The particle acceleration mechanisms in blazar jets are fundamental to high-energy astrophysics. However, the classification of these mechanisms among blazar subclasses and their intrinsic physical characteristics remain elusive (\citealt{Xiao_2024}). The significant flux state in TS $\geq 4$ observed in monthly time bins, relative to the surrounding bins characterized by upper limits, is indicative evidence for the acceleration of particles that lead to the generation of these significant $\gamma$-rays through either leptonic and hadronic mechanisms. The distinguishing features of these leptonic and hadronic processes are outlined in \cite{MUCKE_2003}.
\section{Conclusions}
In this study, we used randomly sampled events based on the distribution of real observations of high-energy neutrino events to search for an excess of neutrino events arriving from the directions of \textit{Fermi} LAT blazar sources. We then performed analysis of $\gamma$-ray light curves of blazars to identify plausible sources of high-energy neutrinos that coincided with the sources within an angular separation of three degrees. Additionally, we performed analysis of the spectral characteristics of these blazars, which could be useful to determine if they are plausible sources of high-energy neutrinos.  The key findings of this study are summarized below.\

\begin{enumerate}
\item The total coincidence of 262 neutrino events with the source positions is significantly higher than expected from random fluctuations alone. Our result, with a significance of 2.42$\sigma$, suggests that the observed coincidence in randomly generated events, distributed uniformly in right ascension and non-uniformly in declination, following the distributions obtained from real observations, is unlikely to be purely due to chance.

\item Based on the spatial and temporal coincidence of the neutrino events, the sources 4FGL J1012.3+0629, 4FGL J2118.0+0019, 4FGL J2226.8+0051, and 4FGL J2227.9+0036 are plausible candidates for high-energy neutrino sources. It also reveals that the high-energy neutrino events IC110726A, IC170308A, IC190415A, IC141210A, IC150102A, IC230524A, IC140114A, IC110807A, and IC200523A are likely to have originated from the blazar sources we identified as plausible candidates.

\item In sources  4FGL J0506.9+0323, 4FGL J1016.0+0512, 4FGL J1018.4+0528, 4FGL J2223.3+0102 and 4FGL J2252.6+1245, there is no $\gamma$-ray activity during the neutrino detection time. Neutrino events located within an angular separation of 3 degrees or less from these sources show no temporal association between their arrival times and periods during which significant photon flux occurs in light curve of these sources.

\item Our analysis of  $\gamma$-ray sources shows that the photon and spectral index fluctuates over time, exhibiting both hardening and softening phases. In time episodes where IceCube neutrino events align with significant $\gamma$-ray flux, the spectral index tends to be relatively harder. This suggests a connection between neutrino production and phases of elevated $\gamma$-ray activity with a harder spectral index. Since hadronic interactions produce both neutrinos and high-energy $\gamma$-rays through \(\pi^0\) decay, spectral hardening during these high-flux states suggests efficient proton acceleration, supporting hadronic models as a key mechanism for neutrino production in blazar jets.

\end{enumerate}
 
\begin{acknowledgements}
     RN acknowledges University Grants Commission Nepal for supporting this research (PhD-79/80-S\&T-15).
\end{acknowledgements}
\bibliographystyle{aa}
\bibliography{refs.bib}

\begin{thebibliography}{60}
\expandafter\ifx\csname natexlab\endcsname\relax\def\natexlab#1{#1}\fi

\bibitem[{Aab {et~al.}(2017)Aab, Abreu, Aglietta, Al~Samarai, Albuquerque,
  Allekotte, Almela, Alvarez~Castillo, Alvarez-Muñiz, Anastasi, Anchordoqui,
  Andrada, Andringa, Aramo, Arqueros, Arsene, Asorey, Assis, Aublin, Avila,
  Badescu, Balaceanu, Barbato, Barreira~Luz, Beatty, Becker, Bellido, Berat,
  Bertaina, Bertou, Biermann, Billoir, Biteau, Blaess, Blanco, Blazek, Bleve,
  Boháčová, Boncioli, Bonifazi, Borodai, Botti, Brack, Brancus, Bretz,
  Bridgeman, Briechle, Buchholz, Bueno, Buitink, Buscemi, Caballero-Mora,
  Caccianiga, Cancio, Canfora, Caramete, Caruso, Castellina, Cataldi, Cazon,
  Chavez, Chinellato, Chudoba, Clay, Cobos, Colalillo, Coleman, Collica,
  Coluccia, Conceição, Consolati, Contreras, Cooper, \& Coutu}]{Aab_2017}
Aab, A., Abreu, P., Aglietta, M., {et~al.} 2017, Sci, 357, 1266

\bibitem[{Aartsen {et~al.}(2017{\natexlab{a}})Aartsen, Ackermann, Adams,
  Aguilar, Ahlers, Ahrens, Altmann, Andeen, Anderson, Ansseau, Anton,
  Archinger, Argüelles, Auer, Auffenberg, Axani, Baccus, Bai, Barnet, Barwick,
  Baum, Bay, Beattie, Beatty, Tjus, Becker, Bendfelt, BenZvi, Berley,
  Bernardini, Bernhard, Besson, Binder, Bindig, Bissok, Blaufuss, Blot,
  Boersma, Bohm, Börner, Bos, Bose, Böser, Botner, Bouchta, Braun, Brayeur,
  Bretz, Bron, Burgman, Burreson, Carver, Casier, Cheung, Chirkin, Christov,
  Clark, Classen, Coenders, Collin, Conrad, Cowen, Cross, Day, Day, de~André,
  Clercq, del Pino~Rosendo, Dembinski, Ridder, Descamps, Desiati, de~Vries,
  de~Wasseige, de~With, DeYoung, Díaz-Vélez, di~Lorenzo, Dujmovic, Dumm,
  Dunkman, Eberhardt, Edwards, Ehrhardt, Eichmann, Eller, Euler, Evenson,
  Fahey, Fazely, Feintzeig, Felde, Filimonov, Finley, Flis, Fösig,
  Franckowiak, Frère, Friedman, Fuchs, Gaisser, Gallagher, Gerhardt, Ghorbani,
  Giang, Gladstone, Glauch, Glowacki, Glüsenkamp, Goldschmidt, Gonzalez,
  Grant, Griffith, Gustafsson, Haack, Hallgren, Halzen, Hansen, Hansmann,
  Hanson, Haugen, Hebecker, Heereman, Helbing, Hellauer, Heller, Hickford,
  Hignight, Hill, Hoffman, Hoffmann, Hoshina, Huang, Huber, Hulth, Hultqvist,
  In, Inaba, Ishihara, Jacobi, Jacobsen, Japaridze, Jeong, Jero, Jones, Jones,
  Joseph, Kang, Kappes, Karg, Karle, Katz, Kauer, Keivani, Kelley, Kemp,
  Kheirandish, Kim, Kim, Kintscher, Kiryluk, Kitamura, Kittler, Klein,
  Kleinfelder, Kleist, Kohnen, Koirala, Kolanoski, Konietz, Köpke, Kopper,
  Kopper, Koskinen, Kowalski, Krasberg, Krings, Kroll, Krückl, Krüger,
  Kunnen, Kunwar, Kurahashi, Kuwabara, Labare, Laihem, Landsman, Lanfranchi,
  Larson, Lauber, Laundrie, Lennarz, Leich, Lesiak-Bzdak, Leuermann, Lu,
  Ludwig, Lünemann, Mackenzie, Madsen, Maggi, Mahn, Mancina, Mandelartz,
  Maruyama, Mase, Matis, Maunu, McNally, McParland, Meade, Meagher, Medici,
  Meier, Meli, Menne, Merino, Meures, Miarecki, Minor, Montaruli, Moulai,
  Murray, Nahnhauer, Naumann, Neer, Newcomb, Niederhausen, Nowicki, Nygren,
  Pollmann, Olivas, O'Murchadha, Palczewski, Pandya, Pankova, Patton, Peiffer,
  Penek, Pepper, de~los Heros, Pettersen, Pieloth, Pinat, Price, Przybylski,
  Quinnan, Raab, Rädel, Rameez, Rawlins, Reimann, Relethford, Relich, Resconi,
  Rhode, Richman, Riedel, Robertson, Rongen, Roucelle, Rott, Ruhe, Ryckbosch,
  Rysewyk, Sabbatini, Herrera, Sandrock, Sandroos, Sandstrom, Sarkar,
  Satalecka, Schlunder, Schmidt, Schoenen, Schöneberg, Schukraft, Schumacher,
  Seckel, Seunarine, Solarz, Soldin, Song, Spiczak, Spiering, Stanev, Stasik,
  Stettner, Steuer, Stezelberger, Stokstad, Stößl, Ström, Strotjohann,
  Sulanke, Sullivan, Sutherland, Taavola, Taboada, Tatar, Tenholt,
  Ter-Antonyan, Terliuk, Tešić, Thollander, Tilav, Toale, Tobin, Toscano,
  Tosi, Tselengidou, Turcati, Unger, Usner, Vandenbroucke, van Eijndhoven,
  Vanheule, van Rossem, van Santen, Vehring, Voge, Vogel, Vraeghe, Wahl, Walck,
  Wallace, Wallraff, Wandkowsky, Weaver, Weiss, Wendt, Westerhoff, Wharton,
  Whelan, Wickmann, Wiebe, Wiebusch, Wille, Williams, Wills, Wisniewski, Wolf,
  Wood, Woolsey, Woschnagg, Xu, Xu, Xu, Yanez, Yodh, Yoshida, \&
  Zoll}]{Aartsen_2017c}
Aartsen, M., Ackermann, M., Adams, J., {et~al.} 2017{\natexlab{a}}, JInst, 12,
  P03012

\bibitem[{Aartsen {et~al.}(2018{\natexlab{a}})Aartsen, Ackermann, Adams,
  Aguilar, Ahlers, Ahrens, Al~Samarai, Altmann, Andeen, Anderson, Ansseau,
  Anton, Argüelles, Arsioli, Auffenberg, Axani, Bagherpour, Bai, Barron,
  Barwick, Baum, Bay, Beatty, Becker, Becker~Tjus, BenZvi, Berley, Bernardini,
  Besson, Binder, Bindig, Blaufuss, Blot, Bohm, Boerner, Bos, Boeser, Botner,
  Bourbeau, Bourbeau, Bradascio, Braun, Brenzke, Bretz, Bron, Brostean-Kaiser,
  Burgman, Busse, Carver, Cheung, Chirkin, Christov, Clark, Classen, Coenders,
  Collin, Conrad, Coppin, Correa, Cowen, Cross, Dave, Day, de~André,
  De~Clercq, Delaunay, Dembinski, DeRidder, Desiati, de~Vries, DeWasseige,
  DeWith, DeYoung, Díaz-Vélez, Di~Lorenzo, Dujmovic, Dumm, Dunkman, Dvorak,
  Eberhardt, Ehrhardt, Eichmann, Eller, Evenson, Fahey, Fazely, Felde,
  Filimonov, Finley, Flis, Franckowiak, Friedman, Fritz, Gaisser, Gallagher,
  Gerhardt, Ghorbani, Giommi, Glauch, Gluesenkamp, Goldschmidt, Gonzalez,
  Grant, Griffith, Haack, Hallgren, Halzen, Hanson, Hebecker, Heereman,
  Helbing, Hellauer, Hickford, Hignight, Hill, Hoffman, Hoffmann, Hoinka,
  Hokanson-Fasig, Hoshina, Huang, Huber, Hultqvist, Huennefeld, Hussain, In,
  Iovine, Ishihara, Jacobi, Japaridze, Jeong, Jero, Jones, Kalaczynski, Kang,
  Kappes, Kappesser, Karg, Karle, Katz, Kauer, Keivani, Kelley, Kheirandish,
  Kim, Kim, Kintscher, Kiryluk, Kittler, Klein, Koirala, Kolanoski, Koepke,
  Kopper, Kopper, Koschinsky, Koskinen, Kowalski, Krammer, Krings, Kroll,
  Krueckl, Kunwar, Neilson, Kuwabara, Kyriacou, Labare, Lanfranchi, Larson,
  Lauber, Leonard, Lesiak-Bzdak, Leuermann, Liu, Lozano~Mariscal, Lu,
  Luenemann, Luszczak, Madsen, Maggi, Mahn, Mancina, Maruyama, Mase, Maunu,
  Meagher, Medici, Meier, Menne, Merino, Meures, Miarecki, Micallef, Momente,
  Montaruli, Moore, Morse, Moulai, Nahnhauer, Nakarmi, Naumann, Neer,
  Niederhausen, Nowicki, Nygren, Pollmann, Olivas, Ó~Murchadha, O’Sullivan,
  Padovani, Palczewski, Pandya, Pankova, Peiffer, Pepper, Perez de~los Heros,
  Pieloth, Pinat, Plum, Price, Przybylski, Raab, Raedel, Rameez, Rawlins, Rea,
  Reimann, Relethford, Relich, Resconi, Rhode, Richman, Robertson, Rongen,
  Rott, Ruhe, Ryckbosch, Rysewyk, Safa, Saelzer, Sahakyan, Sanchez~Herrera,
  Sandrock, Sandroos, Santander, Sarkar, Sarkar, Satalecka, Schlunder, Schmidt,
  Schneider, Schoenen, Schoeneberg, Schumacher, Sclanfani, Seckel, Seunarine,
  Soedingrekso, Soldin, Song, Spiczak, Spiering, Stachurska, Stamatikos,
  Stanev, Stasik, Stettner, Steuer, Stezelberger, Stokstad, Stoessl,
  Strotjohann, Stuttard, Sullivan, Sutherland, Taboada, Tatar, Tenholt,
  Ter-Antonyan, Terliuk, Tilav, Toale, Tobin, Toennis, Toscano, Tosi,
  Tselengidou, Tung, Turcati, Turley, Ty, Unger, Usner, Van~Driessche,
  Van~Eijk, van Eijndhoven, Vandenbroucke, Vanheule, van Santen, Vogel,
  Vraeghe, Walck, Wallace, Wallraff, Wandler, Wandkowsky, Waza, Weaver, Weiss,
  Wendt, Werthebach, Westerhoff, Whelan, Whitehorn, Wiebe, Wiebusch, Wille,
  Williams, Wills, Wolf, Wood, Wood, Woschnagg, Xu, Xu, Xu, Yanez, Yodh,
  Yoshida, \& Yuan}]{AartsenM_2018}
Aartsen, M., Ackermann, M., Adams, J., {et~al.} 2018{\natexlab{a}}, Sci, 361,
  147–151

\bibitem[{Aartsen {et~al.}(2018{\natexlab{b}})Aartsen, Ackermann, Adams,
  Aguilar, Ahlers, Ahrens, Al~Samarai, Altmann, Andeen, Anderson, Ansseau,
  Anton, Argüelles, Auffenberg, Axani, Bagherpour, Bai, Barron, Barwick, Baum,
  Bay, Beatty, Becker, Tjus, BenZvi, Berley, Bernardini, Besson, Binder,
  Bindig, Blaufuss, Blot, Bohm, Boerner, Bos, Boeser, Botner, Bourbeau,
  Bourbeau, Bradascio, Braun, Brenzke, Bretz, Bron, Brostean-Kaiser, Burgman,
  Busse, Carver, Cheng, Chirkin, Christov, Clark, Classen, Coenders, Collin,
  Conrad, Coppin, Correa, Cowen, Cross, Dave, Day, de~Andre, Clercq, Delaunay,
  Dembinski, DeRidder, Desiati, de~Vries, DeWasseige, DeWith, DeYoung,
  Díaz-Vélez, Lorenzo, Dujmovic, Dumm, Dunkman, Dvorak, Eberhardt, Ehrhardt,
  Eichmann, Eller, Evenson, Fahey, Fazely, Felde, Filimonov, Finley, Flis,
  Franckowiak, Friedman, Fritz, Gaisser, Gallagher, Gerhardt, Ghorbani, Glauch,
  Gluesenkamp, Goldschmidt, Gonzalez, Grant, Griffith, Haack, Hallgren, Halzen,
  Hanson, Hebecker, Heereman, Helbing, Hellauer, Hickford, Hignight, Hill,
  Hoffman, Hoffmann, Hoinka, Hokanson-Fasig, Hoshina, Huang, Huber, Hultqvist,
  Huennefeld, Hussain, In, Iovine, Ishihara, Jacobi, Japaridze, Jeong, Jero,
  Jones, Kalaczynski, Kang, Kappes, Kappesser, Karg, Karle, Katz, Kauer,
  Keivani, Kelley, Kheirandish, Kim, Kim, Kintscher, Kiryluk, Kittler, Klein,
  Koirala, Kolanoski, Koepke, Kopper, Kopper, Koschinsky, Koskinen, Kowalski,
  Krings, Kroll, Krueckl, Kunwar, Neilson, Kuwabara, Kyriacou, Labare,
  Lanfranchi, Larson, Lauber, Leonard, Lesiak-Bzdak, Leuermann, Liu, Mariscal,
  Lu, Luenemann, Luszczak, Madsen, Maggi, Mahn, Mancina, Maruyama, Mase, Maunu,
  Meagher, Medici, Meier, Menne, Merino, Meures, Miarecki, Micallef, Momente,
  Montaruli, Moore, Morse, Moulai, Nahnhauer, Nakarmi, Naumann, Neer,
  Niederhausen, Nowicki, Nygren, Pollmann, Olivas, Murchadha, O'Sullivan,
  Palczewski, Pandya, Pankova, Peiffer, Pepper, de~los Heros, Pieloth, Pinat,
  Plum, Price, Przybylski, Raab, Raedel, Rameez, Rauch, Rawlins, Rea, Reimann,
  Relethford, Relich, Resconi, Rhode, Richman, Robertson, Rongen, Rott, Ruhe,
  Ryckbosch, Rysewyk, Safa, Saelzer, Sanchez, Sandrock, Sandroos, Santander,
  Sarkar, Sarkar, Satalecka, Schlunder, Schmidt, Schneider, Schoenen,
  Schoneberg, Schumacher, Sclanfani, Seckel, Seunarine, Soedingrekso, Soldin,
  Song, Spiczak, Spiering, Stachurska, Stamatikos, Stanev, Stasik, Stein,
  Stettner, Steuer, Stezelberger, Stokstad, Stoessl, Strotjohann, Stuttard,
  Sullivan, Sutherland, Taboada, Tatar, Tenholt, Ter-Antonyan, Terliuk, Tilav,
  Toale, Tobin, Toennis, Toscano, Tosi, Tselengidou, Tung, Turcati, Turley, Ty,
  Unger, Usner, Driessche, Eijk, van Eijndhoven, Vandenbroucke, Vanheule, van
  Santen, Vogel, Vraeghe, Walck, Wallace, Wallraff, Wandler, Wandkowsky, Waza,
  Weaver, Weiss, Wendt, Werthebach, Westerhoff, Whelan, Whitehorn, Wiebe,
  Wiebusch, Wille, Williams, Wills, Wolf, Wood, Wood, Woschnagg, Xu, Xu, Xu,
  Yanez, Yodh, Yoshida, Yuan, Abdollahi, Ajello, Angioni, Baldini, Ballet,
  Barbiellini, Bastieri, Bechtol, Bellazzini, Berenji, Bissaldi, Blandford,
  Bonino, Bottacini, Bregeon, Bruel, Büehler, Burnett, Burns, Buson, Cameron,
  Caputo, Caraveo, Cavazzuti, Charles, Chen, Cheung, Chiang, Chiaro, Ciprini,
  Cohen-Tanugi, Conrad, Costantin, Cutini, D'Ammando, de~Palma, Digel, Lalla,
  Mauro, Venere, Domínguez, Favuzzi, Franckowiak, Fukazawa, Funk, Fusco,
  Gargano, Gasparrini, Giglietto, Giomi, Giommi, Giordano, Giroletti, Glanzman,
  Green, Grenier, Grondin, Guiriec, Harding, Hayashida, Hays, Hewitt, Horan,
  Jóhannesson, Kadler, Kensei, Kocevski, Krauss, Kreter, Kuss, Mura, Larsson,
  Latronico, Lemoine-Goumard, Li, Longo, Loparco, Lovellette, Lubrano, Magill,
  Maldera, Malyshev, Manfreda, Mazziotta, McEnery, Meyer, Michelson, Mizuno,
  Monzani, Morselli, Moskalenko, Negro, Nuss, Ojha, Omodei, Orienti, Orlando,
  Palatiello, Paliya, Perkins, Persic, Pesce-Rollins, Piron, Porter, Principe,
  Rainò, Rando, Rani, Razzano, Razzaque, Reimer, Reimer, Renault-Tinacci,
  Ritz, Rochester, Parkinson, Sgrò, Siskind, Spandre, Spinelli, Suson, Tajima,
  Takahashi, Tanaka, Thayer, Thompson, Tibaldo, Torres, Torresi, Tosti, Troja,
  Valverde, Vianello, Vogel, Wood, Wood, Zaharijas, Ahnen, Ansoldi, Antonelli,
  Arcaro, Baack, Babić, Banerjee, Bangale, de~Almeida, Barrio, González,
  Bednarek, Bernardini, Berti, Bhattacharyya, Biland, Blanch, Bonnoli, Carosi,
  Carosi, Ceribella, Chatterjee, Colak, Colin, Colombo, Contreras, Cortina,
  Covino, Cumani, Vela, Dazzi, Angelis, Lotto, Delfino, Delgado, Pierro,
  Domínguez, Prester, Dorner, Doro, Einecke, Elsaesser, Ramazani,
  Fernández-Barral, Fidalgo, Foffano, Pfrang, Fonseca, Font, Fruck, Galindo,
  Gallozzi, López, Garczarczyk, Gaug, Giammaria, Godinović, Gora, Guberman,
  Hadasch, Hahn, Hassan, Hayashida, Herrera, Hose, Hrupec, Inoue, Ishio, Konno,
  Kubo, Kushida, Lelas, Lindfors, Lombardi, Longo, López, Maggio, Majumdar,
  Makariev, Maneva, Manganaro, Mannheim, Maraschi, Mariotti, Martínez, Masuda,
  Mazin, Minev, Miranda, Mirzoyan, Moralejo, Moreno, Moretti, Nagayoshi,
  Neustroev, Niedzwiecki, Rosillo, Nigro, Nilsson, Ninci, Nishijima, Noda,
  Nogués, Paiano, Palacio, Paneque, Paoletti, Paredes, Pedaletti, Peresano,
  Persic, Moroni, Prandini, Puljak, Rodriguez, Reichardt, Rhode, Ribó, Rico,
  Righi, Rugliancich, Saito, Satalecka, Schweizer, Sitarek, Šnidarić,
  Sobczynska, Stamerra, Strzys, Surić, Takahashi, Tavecchio, Temnikov,
  Terzić, Teshima, Torres-Albà, Treves, Tsujimoto, Vanzo, Acosta, Vovk, Ward,
  Will, Zarić, Franceschini, Lucarelli, Tavani, Piano, Donnarumma, Pittori,
  Verrecchia, Barbiellini, Bulgarelli, Caraveo, Cattaneo, Colafrancesco, Costa,
  Cocco, Ferrari, Gianotti, Giuliani, Lipari, Mereghetti, Morselli, Pacciani,
  Paoletti, Parmiggiani, Pellizzoni, Picozza, Pilia, Rappoldi, Trois,
  Vercellone, Vittorini, Albert, Alfaro, Álvarez, Arceo, Velázquez, Rojas,
  Solares, Becerril, Belmont-Moreno, Bernal, Mora, Rojas, Carramiñana,
  Casanova, Maldonado, Cotti, Cotzomi, de~León, Acuña, la~Fuente, Hernandez,
  Dichiara, Dingus, DuVernois, Velez, Ellsworth, Engel, Fiorino, Fleischhack,
  Fraija, González, Garfias, González, Muñoz, Goodman, Hampel-Arias,
  Harding, Cadena, Hona, Hueyotl-Zahuantitla, Hui, Hüntemeyer, Iriarte,
  Jardin-Blicq, Joshi, Kaufmann, Kunde, Lara, Lauer, Lee, Lennarz, Vargas,
  Linnemann, Longinotti, Luis-Raya, Luna-García, Malone, Marinelli, Martinez,
  Castellanos, Huerta, Castro, Matthews, Miranda-Romagnoli, Barbosa, Mostafa,
  Nayerhoda, Nellen, Newbold, Nisa, Noriega-Papaqui, Pelayo, Pretz, Pérez,
  Ren, Rho, Rivière, González, Rosenberg, Ruiz-Velasco, Ruiz-Velasco, Greus,
  Sandoval, Schneider, Schoorlemmer, Sinnis, Smith, Springer, Surajbali,
  Tibolla, Tollefson, Torres, Villaseñor, Weisgarber, Werner, Yapici, Yodh,
  Zepeda, Zhou, de~Dios~Álvarez Romero, Abdalla, Angüner, Armand, Backes,
  Becherini, Berge, Böttcher, Boisson, Bolmont, Bonnefoy, Bordas, Brun,
  Büchele, Bulik, Caroff, Carosi, Casanova, Cerruti, Chakraborty, Chandra,
  Chen, Colafrancesco, Davids, Deil, Devin, Djannati-Ataï, Egberts, Emery,
  Eschbach, Fiasson, Fontaine, Funk, Füßling, Gallant, Gaté, Giavitto,
  Glawion, Glicenstein, Gottschall, Grondin, Haupt, Henri, Hinton, Hoischen,
  Holch, Huber, Jamrozy, Jankowsky, Jankowsky, Jouvin, Jung-Richardt,
  Kerszberg, Khélifi, King, Klepser, Kluźniak, Komin, Kraus, Lefaucheur,
  Lemière, Lemoine-Goumard, Lenain, Leser, Lohse, López-Coto, Lorentz,
  Lypova, Marandon, Martí-Devesa, Maurin, Mitchell, Moderski, Mohamed,
  Mohrmann, Moulin, Murach, de~Naurois, Niederwanger, Niemiec, Oakes, O'Brien,
  Ohm, Ostrowski, Oya, Panter, Parsons, Perennes, Piel, Pita, Poireau, Noel,
  Prokoph, Pühlhofer, Quirrenbach, Raab, Rauth, Renaud, Rieger, Rinchiuso,
  Romoli, Rowell, Rudak, Sanchez, Sasaki, Schlickeiser, Schüssler, Schulz,
  Schwanke, Seglar-Arroyo, Shafi, Simoni, Sol, Stegmann, Steppa, Tavernier,
  Taylor, Tiziani, Trichard, Tsirou, van Eldik, van Rensburg, van Soelen, Veh,
  Vincent, Voisin, Wagner, Wagner, Wierzcholska, Zanin, Zdziarski, Zech,
  Ziegler, Zorn, Zywucka, Savchenko, Ferrigno, Bazzano, Diehl, Kuulkers,
  Laurent, Mereghetti, Natalucci, Panessa, Rodi, Ubertini, Morokuma, Ohta,
  Tanaka, Mori, Yamanaka, Kawabata, Utsumi, Nakaoka, Kawabata, Nagashima,
  Yoshida, Matsuoka, Itoh, Keel, Copperwheat, Steele, Cenko, Evans, Fox,
  Kennea, Marshall, Osborne, Tohuvavohu, Turley, Cowen, DeLaunay, Keivani,
  Santander, Abeysekara, Archer, Benbow, Bird, Brill, Brose, Buchovecky,
  Buckley, Bugaev, Christiansen, Connolly, Cui, Daniel, Errando, Falcone, Feng,
  Finley, Fortson, Furniss, Gueta, Hütten, Hervet, Hughes, Humensky, Johnson,
  Kaaret, Kar, Kelley-Hoskins, Kertzman, Kieda, Krause, Krennrich, Kumar, Lang,
  Lin, Maier, McArthur, Moriarty, Mukherjee, Nieto, O'Brien, Ong, Otte, Park,
  Petrashyk, Pohl, Popkow, Pueschel, Quinn, Ragan, Reynolds, Richards, Roache,
  Rulten, Sadeh, Santander, Scott, Sembroski, Shahinyan, Sushch, Trépanier,
  Tyler, Vassiliev, Wakely, Weinstein, Wells, Wilcox, Wilhelm, Williams,
  Zitzer, Tetarenko, Kimball, Miller-Jones, \& Sivakoff}]{Aartsen_2018a}
Aartsen, M., Ackermann, M., Adams, J., {et~al.} 2018{\natexlab{b}}, Sci, 361,
  eaat1378

\bibitem[{Aartsen {et~al.}(2013)Aartsen, Abbasi, Abdou, Ackermann, Adams,
  Aguilar, Ahlers, Altmann, Auffenberg, Bai, Baker, Barwick, Baum, Bay, Beatty,
  Bechet, Becker~Tjus, Becker, Benabderrahmane, BenZvi, Berghaus, Berley,
  Bernardini, Bernhard, Bertrand, Besson, Binder, Bindig, Bissok, Blaufuss,
  Blumenthal, Boersma, Bohaichuk, Bohm, Bose, Böser, Botner, Brayeur, Bretz,
  Brown, Bruijn, Brunner, Carson, Casey, Casier, Chirkin, Christov, Christy,
  Clark, Clevermann, Coenders, Cohen, Cowen, Cruz~Silva, Danninger, Daughhetee,
  Davis, Day, De~Clercq, De~Ridder, Desiati, de~Vries, de~With, DeYoung,
  Díaz-Vélez, Dunkman, Eagan, Eberhardt, Eichmann, Eisch, Ellsworth, Euler,
  Evenson, Fadiran, Fazely, Fedynitch, Feintzeig, Feusels, Filimonov, Finley,
  Fischer-Wasels, Flis, Franckowiak, Frantzen, Fuchs, Gaisser, Gallagher,
  Gerhardt, Gladstone, Glüsenkamp, Goldschmidt, Golup, Gonzalez, Goodman,
  Góra, Grandmont, Grant, Groß, Ha, Haj~Ismail, Hallen, Hallgren, Halzen,
  Hanson, Heereman, Heinen, Helbing, Hellauer, Hickford, Hill, Hoffman,
  Hoffmann, Homeier, Hoshina, Huelsnitz, Hulth, Hultqvist, Hussain, Ishihara,
  Jacobi, Jacobsen, Jagielski, Japaridze, Jero, Jlelati, Kaminsky, Kappes,
  Karg, Karle, Kelley, Kiryluk, Kläs, Klein, Köhne, Kohnen, Kolanoski,
  Köpke, Kopper, Kopper, Koskinen, Kowalski, Krasberg, Krings, Kroll, Kunnen,
  Kurahashi, Kuwabara, Labare, Landsman, Larson, Lesiak-Bzdak, Leuermann,
  Leute, Lünemann, Madsen, Maggi, Maruyama, Mase, Matis, McNally, Meagher,
  Merck, Meures, Miarecki, Middell, Milke, Miller, Mohrmann, Montaruli, Morse,
  Nahnhauer, Naumann, Niederhausen, Nowicki, Nygren, \&
  Obertacke}]{Aartsen_2013}
Aartsen, M.~G., Abbasi, R., Abdou, Y., {et~al.} 2013, Science, 342, 1242856

\bibitem[{Aartsen {et~al.}(2017{\natexlab{b}})Aartsen, Ackermann, Adams,
  Aguilar, Ahlers, Ahrens, Samarai, Altmann, Andeen, Anderson, Ansseau, Anton,
  Argüelles, Auffenberg, Axani, Bagherpour, Bai, Barwick, Baum, Bay, Beatty,
  Tjus, Becker, BenZvi, Berley, Bernardini, Besson, Binder, Bindig, Blaufuss,
  Blot, Bohm, Börner, Bos, Bose, Böser, Botner, Bourbeau, Bradascio, Braun,
  Brayeur, Brenzke, Bretz, Bron, Burgman, Carver, Casey, Casier, Cheung,
  Chirkin, Christov, Clark, Classen, Coenders, Collin, Conrad, Cowen, Cross,
  Day, de~André, Clercq, DeLaunay, Dembinski, Ridder, Desiati, de~Vries,
  de~Wasseige, de~With, DeYoung, Díaz-Vélez, di~Lorenzo, Dujmovic, Dumm,
  Dunkman, Eberhardt, Ehrhardt, Eichmann, Eller, Evenson, Fahey, Fazely, Felde,
  Filimonov, Finley, Flis, Franckowiak, Friedman, Fuchs, Gaisser, Gallagher,
  Gerhardt, Ghorbani, Giang, Glauch, Glüsenkamp, Goldschmidt, Gonzalez, Grant,
  Griffith, Haack, Hallgren, Halzen, Hanson, Hebecker, Heereman, Helbing,
  Hellauer, Hickford, Hignight, Hill, Hoffman, Hoffmann, Hokanson-Fasig,
  Hoshina, Huang, Huber, Hultqvist, In, Ishihara, Jacobi, Japaridze, Jeong,
  Jero, Jones, Kalacynski, Kang, Kappes, Karg, Karle, Katz, Kauer, Keivani,
  Kelley, Kheirandish, Kim, Kim, Kintscher, Kiryluk, Kittler, Klein, Kohnen,
  Koirala, Kolanoski, Köpke, Kopper, Kopper, Koschinsky, Koskinen, Kowalski,
  Krings, Kroll, Krückl, Kunnen, Kunwar, Kurahashi, Kuwabara, Kyriacou,
  Labare, Lanfranchi, Larson, Lauber, Lennarz, Lesiak-Bzdak, Leuermann, Liu,
  Lu, Lünemann, Luszczak, Madsen, Maggi, Mahn, Mancina, Maruyama, Mase, Maunu,
  McNally, Meagher, Medici, Meier, Menne, Merino, Meures, Miarecki, Micallef,
  Momenté, Montaruli, Moulai, Nahnhauer, Nakarmi, Naumann, Neer, Niederhausen,
  Nowicki, Nygren, Pollmann, Olivas, O’Murchadha, Palczewski, Pandya,
  Pankova, Peiffer, Pepper, de~los Heros, Pieloth, Pinat, Plum, Price,
  Przybylski, Raab, Rädel, Rameez, Rawlins, Reimann, Relethford, Relich,
  Resconi, Rhode, Richman, Riedel, Robertson, Rongen, Rott, Ruhe, Ryckbosch,
  Rysewyk, Sälzer, Herrera, Sandrock, Sandroos, Sarkar, Sarkar, Satalecka,
  Schlunder, Schmidt, Schneider, Schoenen, Schöneberg, Schumacher, Seckel,
  Seunarine, Soldin, Song, Spiczak, Spiering, Stachurska, Stanev, Stasik,
  Stettner, Steuer, Stezelberger, Stokstad, Stößl, Strotjohann, Sullivan,
  Sutherland, Taboada, Tatar, Tenholt, Ter-Antonyan, Terliuk, Tešić, Tilav,
  Toale, Tobin, Toscano, Tosi, Tselengidou, Tung, Turcati, Turley, Ty, Unger,
  Usner, Vandenbroucke, Driessche, van Eijndhoven, Vanheule, van Santen,
  Vehring, Vogel, Vraeghe, Walck, Wallace, Wallraff, Wandkowsky, Waza, Weaver,
  Weiss, Wendt, Westerhoff, Whelan, Wickmann, Wiebe, Wiebusch, Wille, Williams,
  Wills, Wolf, Wood, Wood, Woolsey, Woschnagg, Xu, Xu, Xu, Yanez, Yodh,
  Yoshida, Yuan, Zoll, \& Collaboration}]{Aartsen_2017}
Aartsen, M.~G., Ackermann, M., Adams, J., {et~al.} 2017{\natexlab{b}}, ApJ,
  846, 136

\bibitem[{Abbasi {et~al.}(2023)Abbasi, Ackermann, Adams, Agarwalla, Aguilar,
  Ahlers, Alameddine, Amin, Andeen, Anton, Argüelles, Ashida, Athanasiadou,
  Axani, Bai, V, Baricevic, Barwick, Basu, Bay, Beatty, Becker, Tjus, Beise,
  Bellenghi, BenZvi, Berley, Bernardini, Besson, Binder, Bindig, Blaufuss,
  Blot, Bontempo, Book, Meneguolo, Böser, Botner, Böttcher, Bourbeau, Braun,
  Brinson, Brostean-Kaiser, Burley, Busse, Butterfield, Campana, Carloni,
  Carnie-Bronca, Chattopadhyay, Chau, Chen, Chen, Chirkin, Choi, Clark,
  Classen, Coleman, Collin, Connolly, Conrad, Coppin, Correa, Countryman,
  Cowen, Dave, Clercq, DeLaunay, Delgado, Dembinski, Deng, Deoskar, Desai,
  Desiati, de~Vries, de~Wasseige, DeYoung, Diaz, Díaz-Vélez, Dittmer, Domi,
  Dujmovic, DuVernois, Ehrhardt, Eller, Engel, Erpenbeck, Evans, Evenson, Fan,
  Fang, Farrag, Fazely, Fedynitch, Feigl, Fiedlschuster, Finley, Fischer, Fox,
  Franckowiak, Friedman, Fritz, Fürst, Gaisser, Gallagher, Ganster, Garcia,
  Gerhardt, Ghadimi, Glaser, Glauch, Glüsenkamp, Goehlke, Gonzalez, Goswami,
  Grant, Gray, Griffin, Griswold, Günther, Gutjahr, Haack, Hallgren, Halliday,
  Halve, Halzen, Hamdaoui, Minh, Hanson, Hardin, Harnisch, Hatch, Haungs,
  Helbing, Hellrung, Henningsen, Heuermann, Heyer, Hickford, Hidvegi, Hill,
  Hill, Hoffman, Hoshina, Hou, Huber, Hultqvist, Hünnefeld, Hussain, Hymon,
  In, Ishihara, Jacquart, Janik, Jansson, Japaridze, Jayakumar, Jeong, Jin,
  Jones, Kang, Kang, Kang, Kappes, Kappesser, Kardum, Karg, Karl, Karle, Katz,
  Kauer, Kelley, Zathul, Kheirandish, Kiryluk, Klein, Kochocki, Koirala,
  Kolanoski, Kontrimas, Köpke, Kopper, Koskinen, Koundal, Kovacevich,
  Kowalski, Kozynets, Kruiswijk, Krupczak, Kumar, Kun, Kurahashi, Lad, Gualda,
  Lamoureux, Larson, Lauber, Lazar, Lee, DeHolton, Leszczyńska, Lincetto, Liu,
  Liubarska, Lohfink, Love, Mariscal, Lu, Lucarelli, Ludwig, Luszczak, Lyu,
  Madsen, Mahn, Makino, Manao, Mancina, Sainte, Mariş, Marka, Marka, Marsee,
  Martinez-Soler, Maruyama, Mayhew, McElroy, McNally, Mead, Meagher, Mechbal,
  Medina, Meier, Merckx, Merten, Micallef, Montaruli, Moore, Morii, Morse,
  Moulai, Mukherjee, Naab, Nagai, Nakos, Naumann, Necker, Neumann,
  Niederhausen, Nisa, Noell, Nowicki, Pollmann, O’Dell, Oehler, Oeyen,
  Olivas, Orsoe, Osborn, O’Sullivan, Pandya, Park, Parker, Paudel, Paul,
  de~los Heros, Peterson, Philippen, Pieper, Pizzuto, Plum, Pontén, Popovych,
  Rodriguez, Pries, Procter-Murphy, Przybylski, Rack-Helleis, Rawlins, Rechav,
  Rehman, Reichherzer, Renzi, Resconi, Reusch, Rhode, Richman, Riedel, Roberts,
  Robertson, Rodan, Roellinghoff, Rongen, Rott, Ruhe, Ruohan, Ryckbosch, Safa,
  Saffer, Salazar-Gallegos, Sampathkumar, Herrera, Sandrock, Santander, Sarkar,
  Sarkar, Savelberg, Savina, Schaufel, Schieler, Schindler, Schlüter,
  Schlüter, Schmidt, Schneider, Schröder, Schumacher, Schwefer, Sclafani,
  Seckel, Seunarine, Shah, Sharma, Shefali, Shimizu, Silva, Skrzypek, Smithers,
  Snihur, Soedingrekso, Søgaard, Soldin, Sommani, Spannfellner, Spiczak,
  Spiering, Stamatikos, Stanev, Stezelberger, Stürwald, Stuttard, Sullivan,
  Taboada, Ter-Antonyan, Thiesmeyer, Thompson, Thwaites, Tilav, Tollefson,
  Tönnis, Toscano, Tosi, Trettin, Tung, Turcotte, Twagirayezu, Ty, Elorrieta,
  Upadhyay, Upshaw, Valtonen-Mattila, Vandenbroucke, van Eijndhoven, Vannerom,
  van Santen, Vara, Veitch-Michaelis, Venugopal, Verpoest, Veske, Walck,
  Watson, Weaver, Weigel, Weindl, Weldert, Wendt, Werthebach, Weyrauch,
  Whitehorn, Wiebusch, Willey, Williams, Wolf, Wolf, Wrede, Xu, Yanez,
  Yildizci, Yoshida, Yu, Yu, Yuan, Zhang, \& Zhelnin}]{Abbasi_2023}
Abbasi, R., Ackermann, M., Adams, J., {et~al.} 2023, ApJS, 269, 25

\bibitem[{Abdo {et~al.}(2010)Abdo, Ackermann, Ajello, Antolini, Baldini,
  Ballet, Barbiellini, Bastieri, Bechtol, Bellazzini, Berenji, Blandford,
  Bloom, Bonamente, Borgland, Bouvier, Bregeon, Brez, Brigida, Bruel, Buehler,
  Burnett, Buson, Caliandro, Cameron, Caraveo, Carrigan, Casandjian, Cavazzuti,
  Cecchi, Çelik, Chekhtman, Cheung, Chiang, Ciprini, Claus, Cohen-Tanugi,
  Cominsky, Conrad, Costamante, Cutini, Dermer, de~Angelis, de~Palma,
  do~Couto~e Silva, Drell, Dubois, Dumora, Farnier, Favuzzi, Fegan, Focke,
  Fortin, Frailis, Fukazawa, Funk, Fusco, Gargano, Gasparrini, Gehrels,
  Germani, Giebels, Giglietto, Giommi, Giordano, Glanzman, Godfrey, Grenier,
  Grondin, Grove, Guiriec, Hadasch, Hayashida, Hays, Healey, Horan, Hughes,
  Itoh, Jóhannesson, Johnson, Johnson, Kamae, Katagiri, Kataoka, Kawai,
  Knödlseder, Kuss, Lande, Larsson, Latronico, Lemoine-Goumard, Longo,
  Loparco, Lott, Lovellette, Lubrano, Madejski, Makeev, Massaro, Mazziotta,
  McEnery, Michelson, Mitthumsiri, Mizuno, Moiseev, Monte, Monzani, Morselli,
  Moskalenko, Mueller, Murgia, Nolan, Norris, Nuss, Ohno, Ohsugi, Omodei,
  Orlando, Ormes, Ozaki, Panetta, Parent, Pelassa, Pepe, Pesce-Rollins, Piron,
  Porter, Rainò, Rando, Razzano, Reimer, Reimer, Ritz, Rodriguez, Romani,
  Roth, Ryde, Sadrozinski, Sander, Scargle, Sgrò, Shaw, Smith, Spandre,
  Spinelli, Starck, Strickman, Suson, Takahashi, Takahashi, Tanaka, Thayer,
  Thayer, Thompson, Tibaldo, Torres, Tosti, Tramacere, Uchiyama, Usher,
  Vasileiou, Vilchez, Vitale, Waite, Wallace, Wang, Winer, Wood, Yang, Ylinen,
  \& Ziegler}]{Abdo_2010b}
Abdo, A.~A., Ackermann, M., Ajello, M., {et~al.} 2010, ApJ, 722, 520

\bibitem[{Abdollahi {et~al.}(2020)Abdollahi, Acero, Ackermann, Ajello, Atwood,
  Axelsson, Baldini, Ballet, Barbiellini, Bastieri, Becerra~Gonzalez,
  Bellazzini, Berretta, Bissaldi, Blandford, Bloom, Bonino, Bottacini, Brandt,
  Bregeon, Bruel, Buehler, Burnett, Buson, Cameron, Caputo, Caraveo,
  Casandjian, Castro, Cavazzuti, Charles, Chaty, Chen, Cheung, Chiaro, Ciprini,
  Cohen-Tanugi, Cominsky, Coronado-Blázquez, Costantin, Cuoco, Cutini,
  D’Ammando, DeKlotz, Torre~Luque, de~Palma, Desai, Digel, Lalla, Mauro,
  Venere, Domínguez, Dumora, Dirirsa, Fegan, Ferrara, Franckowiak, Fukazawa,
  Funk, Fusco, Gargano, Gasparrini, Giglietto, Giommi, Giordano, Giroletti,
  Glanzman, Green, Grenier, Griffin, Grondin, Grove, Guiriec, Harding, Hayashi,
  Hays, Hewitt, Horan, Jóhannesson, Johnson, Kamae, Kerr, Kocevski,
  Kovac’evic’, Kuss, Landriu, Larsson, Latronico, Lemoine-Goumard, Li,
  Liodakis, Longo, Loparco, Lott, Lovellette, Lubrano, Madejski, Maldera,
  Malyshev, Manfreda, Marchesini, Marcotulli, Martí-Devesa, Martin, Massaro,
  Mazziotta, McEnery, Mereu, Meyer, Michelson, Mirabal, Mizuno, Monzani,
  Morselli, Moskalenko, Negro, Nuss, Ojha, Omodei, Orienti, Orlando, Ormes,
  Palatiello, Paliya, Paneque, Pei, Peña-Herazo, Perkins, Persic,
  Pesce-Rollins, Petrosian, Petrov, Piron, Poon, Porter, Principe, Rainò,
  Rando, Razzano, Razzaque, Reimer, Reimer, Remy, Reposeur, Romani, Parkinson,
  Schinzel, Serini, Sgrò, Siskind, Smith, Spandre, Spinelli, Strong, Suson,
  Tajima, Takahashi, Tak, Thayer, Thompson, Tibaldo, Torres, Torresi, Valverde,
  Klaveren, Zyl, Wood, Yassine, \& Zaharijas}]{Abdollahi_2020}
Abdollahi, S., Acero, F., Ackermann, M., {et~al.} 2020, ApJS, 247, 33

\bibitem[{Aharonian {et~al.}(2009)Aharonian, {Akhperjanian, A. G.}, {Anton,
  G.}, {Barres de Almeida, U.}, {Bazer-Bachi, A. R.}, {Becherini, Y.}, {Behera,
  B.}, {Benbow, W.}, {Bernlöhr, K.}, {Boisson, C.}, {Bochow, A.}, {Borrel,
  V.}, {Brion, E.}, {Brucker, J.}, {Brun, P.}, {Bühler, R.}, {Bulik, T.},
  {Büsching, I.}, {Boutelier, T.}, {Chadwick, P. M.}, {Charbonnier, A.},
  {Chaves, R. C. G.}, {Cheesebrough, A.}, {Chounet, L.-M.}, {Clapson, A. C.},
  {Coignet, G.}, {Costamante, L.}, {Dalton, M.}, {Daniel, M. K.}, {Davids, I.
  D.}, {Degrange, B.}, {Deil, C.}, {Dickinson, H. J.}, {Djannati-Ataï, A.},
  {Domainko, W.}, {Drury, L. O'C.}, {Dubois, F.}, {Dubus, G.}, {Dyks, J.},
  {Dyrda, M.}, {Egberts, K.}, {Emmanoulopoulos, D.}, {Espigat, P.}, {Farnier,
  C.}, {Feinstein, F.}, {Fiasson, A.}, {Förster, A.}, {Fontaine, G.},
  {Füßling, M.}, {Gabici, S.}, {Gallant, Y. A.}, {Gérard, L.}, {Giebels,
  B.}, {Glicenstein, J. F.}, {Glück, B.}, {Goret, P.}, {Göhring, D.},
  {Hauser, D.}, {Hauser, M.}, {Heinz, S.}, {Heinzelmann, G.}, {Henri, G.},
  {Hermann, G.}, {Hinton, J. A.}, {Hoffmann, A.}, {Hofmann, W.}, {Holleran,
  M.}, {Hoppe, S.}, {Horns, D.}, {Jacholkowska, A.}, {de Jager, O. C.}, {Jahn,
  C.}, {Jung, I.}, {Katarzyński, K.}, {Katz, U.}, {Kaufmann, S.}, {Kendziorra,
  E.}, {Kerschhaggl, M.}, {Khangulyan, D.}, {Khélifi, B.}, {Keogh, D.},
  {Kluźniak, W.}, {Kneiske, T.}, {Komin, Nu.}, {Kosack, K.}, {Lamanna, G.},
  {Lenain, J.-P.}, {Lohse, T.}, {Marandon, V.}, {Martin, J. M.},
  {Martineau-Huynh, O.}, {Marcowith, A.}, {Maurin, D.}, {McComb, T. J. L.},
  {Medina, M. C.}, {Moderski, R.}, {Monard, L. A. G.}, {Moulin, E.},
  {Naumann-Godo, M.}, {de Naurois, M.}, {Nedbal, D.}, {Nekrassov, D.},
  {Niemiec, J.}, {Nolan, S. J.}, {Ohm, S.}, {Olive, J.-F.}, {de Oña Wilhelmi,
  E.}, {Orford, K. J.}, {Ostrowski, M.}, {Panter, M.}, {Paz Arribas, M.},
  {Pedaletti, G.}, {Pelletier, G.}, {Petrucci, P.-O.}, {Pita, S.}, {Pühlhofer,
  G.}, {Punch, M.}, {Quirrenbach, A.}, {Raubenheimer, B. C.}, {Raue, M.},
  {Rayner, S. M.}, {Renaud, M.}, {Rieger, F.}, {Ripken, J.}, {Rob, L.},
  {Rosier-Lees, S.}, {Rowell, G.}, {Rudak, B.}, {Rulten, C. B.}, {Ruppel, J.},
  {Sahakian, V.}, {Santangelo, A.}, {Schlickeiser, R.}, {Schöck, F. M.},
  {Schröder, R.}, {Schwanke, U.}, {Schwarzburg, S.}, {Schwemmer, S.},
  {Shalchi, A.}, {Sikora, M.}, {Skilton, J. L.}, {Sol, H.}, {Spangler, D.},
  {Stawarz, Ł.}, {Steenkamp, R.}, {Stegmann, C.}, {Superina, G.}, {Szostek,
  A.}, {Tam, P. H.}, {Tavernet, J.-P.}, {Terrier, R.}, {Tibolla, O.},
  {Tluczykont, M.}, {van Eldik, C.}, {Vasileiadis, G.}, {Venter, C.}, {Venter,
  L.}, {Vialle, J. P.}, {Vincent, P.}, {Vivier, M.}, {Völk, H. J.}, {Volpe,
  F.}, {Wagner, S. J.}, {Ward, M.}, {Zdziarski, A. A.}, \& {Zech,
  A.}}]{Aharonian_2009}
Aharonian, F., {Akhperjanian, A. G.}, {Anton, G.}, {et~al.} 2009, A\&A, 502,
  749

\bibitem[{Atoyan \& Dermer(2001)}]{Atoyan_2001}
Atoyan, A. \& Dermer, C.~D. 2001, Phys. Rev. Lett., 87, 221102

\bibitem[{{Atoyan, A. M.} {et~al.}(2002){Atoyan, A. M.}, {Aye, K.-M.},
  {Chadwick, P. M.}, {Daniel, M. K.}, {Lyons, K.}, {McComb, T. J. L.},
  {McKenny, J. M.}, {Nolan, S. J.}, {Orford, K. J.}, {Osborne, J. L.}, \&
  {Rayner, S. M.}}]{Atoyan_2002}
{Atoyan, A. M.}, {Aye, K.-M.}, {Chadwick, P. M.}, {et~al.} 2002, A\&A, 383, 864

\bibitem[{Atwood {et~al.}(2009)Atwood, Abdo, Ackermann, Althouse, Anderson,
  Axelsson, Baldini, Ballet, Band, Barbiellini, Bartelt, Bastieri, Baughman,
  Bechtol, Bédérède, Bellardi, Bellazzini, Berenji, Bignami, Bisello,
  Bissaldi, Blandford, Bloom, Bogart, Bonamente, Bonnell, Borgland, Bouvier,
  Bregeon, Brez, Brigida, Bruel, Burnett, Busetto, Caliandro, Cameron, Caraveo,
  Carius, Carlson, Casandjian, Cavazzuti, Ceccanti, Cecchi, Charles, Chekhtman,
  Cheung, Chiang, Chipaux, Cillis, Ciprini, Claus, Cohen-Tanugi, Condamoor,
  Conrad, Corbet, Corucci, Costamante, Cutini, Davis, Decotigny, DeKlotz,
  Dermer, de~Angelis, Digel, do~Couto~e Silva, Drell, Dubois, Dumora, Edmonds,
  Fabiani, Farnier, Favuzzi, Flath, Fleury, Focke, Funk, Fusco, Gargano,
  Gasparrini, Gehrels, Gentit, Germani, Giebels, Giglietto, Giommi, Giordano,
  Glanzman, Godfrey, Grenier, Grondin, Grove, Guillemot, Guiriec, Haller,
  Harding, Hart, Hays, Healey, Hirayama, Hjalmarsdotter, Horn, Hughes,
  Jóhannesson, Johansson, Johnson, Johnson, Johnson, Johnson, Kamae, Katagiri,
  Kataoka, Kavelaars, Kawai, Kelly, Kerr, Klamra, Knödlseder, Kocian, Komin,
  Kuehn, Kuss, Landriu, Latronico, Lee, Lee, Lemoine-Goumard, Lionetto, Longo,
  Loparco, Lott, Lovellette, Lubrano, Madejski, Makeev, Marangelli, Massai,
  Mazziotta, McEnery, Menon, Meurer, Michelson, Minuti, Mirizzi, Mitthumsiri,
  Mizuno, Moiseev, Monte, Monzani, Moretti, Morselli, Moskalenko, Murgia,
  Nakamori, Nishino, Nolan, Norris, Nuss, Ohno, Ohsugi, Omodei, Orlando, Ormes,
  Paccagnella, Paneque, Panetta, Parent, Pearce, Pepe, Perazzo, Pesce-Rollins,
  Picozza, Pieri, Pinchera, Piron, Porter, Poupard, Rainò, Rando, Rapposelli,
  Razzano, Reimer, Reimer, Reposeur, Reyes, Ritz, Rochester, Rodriguez, Romani,
  Roth, Russell, Ryde, Sabatini, Sadrozinski, Sanchez, Sander, Sapozhnikov,
  Parkinson, Scargle, Schalk, Scolieri, Sgrò, Share, Shaw, Shimokawabe,
  Shrader, Sierpowska-Bartosik, Siskind, Smith, Smith, Spandre, Spinelli,
  Starck, Stephens, Strickman, Strong, Suson, Tajima, Takahashi, Takahashi,
  Tanaka, Tenze, Tether, Thayer, Thayer, Thompson, Tibaldo, Tibolla, Torres,
  Tosti, Tramacere, Turri, Usher, Vilchez, Vitale, Wang, Watters, Winer, Wood,
  Ylinen, \& Ziegler}]{Atwood_2009}
Atwood, W.~B., Abdo, A.~A., Ackermann, M., {et~al.} 2009, ApJ, 697, 1071

\bibitem[{Bhatta \& Dhital(2020)}]{Bhatta_Dhital_2020}
Bhatta, G. \& Dhital, N. 2020, ApJ, 891, 120

\bibitem[{Bhatta {et~al.}(2018)Bhatta, Mohorian, \& Bilinsky}]{Bhatta_2018}
Bhatta, G., Mohorian, M., \& Bilinsky, I. 2018, A\&A, 619, A93

\bibitem[{Bruel {et~al.}(2018)Bruel, Burnett, Digel, Johannesson, Omodei, \&
  Wood}]{bruel2018fermi}
Bruel, P., Burnett, T., Digel, S., {et~al.} 2018, Fermi-LAT improved Pass$\sim$
  8 event selection

\bibitem[{Cerruti {et~al.}(2015)Cerruti, Zech, Boisson, \&
  Inoue}]{Cerruti_2015}
Cerruti, M., Zech, A., Boisson, C., \& Inoue, S. 2015, MNRAS, 448, 910

\bibitem[{Chatterjee {et~al.}(2021)Chatterjee, Roy, Sarkar, \&
  Chitnis}]{Chatterjee_2021}
Chatterjee, A., Roy, A., Sarkar, A., \& Chitnis, V.~R. 2021, MNRAS, 508, 1986

\bibitem[{Das \& Chatterjee(2023)}]{Chatterjee_2023}
Das, S. \& Chatterjee, R. 2023, MNRAS, 524, 3797

\bibitem[{Dermer {et~al.}(2009)Dermer, Razzaque, Finke, \&
  Atoyan}]{Dermer_2009}
Dermer, C.~D., Razzaque, S., Finke, J.~D., \& Atoyan, A. 2009, New Journal of
  Physics, 11, 065016

\bibitem[{Dermer {et~al.}(2015)Dermer, Yan, Zhang, Finke, \&
  Lott}]{Dermer_2015}
Dermer, C.~D., Yan, D., Zhang, L., Finke, J.~D., \& Lott, B. 2015, ApJ, 809,
  174

\bibitem[{Dimitrakoudis {et~al.}(2012)Dimitrakoudis, Mastichiadis, Protheroe,
  \& Reimer}]{Dimitrakoudis_2012}
Dimitrakoudis, S., Mastichiadis, A., Protheroe, R.~J., \& Reimer, A. 2012,
  A\&A, 546, A120

\bibitem[{Doert \& Errando(2014)}]{Doert_2014}
Doert, M. \& Errando, M. 2014, ApJ, 782, 41

\bibitem[{Donnarumma {et~al.}(2009)Donnarumma, Pucella, Vittorini, D'Ammando,
  Vercellone, Raiteri, Villata, Perri, Chen, Smart, Kataoka, Kawai, Mori,
  Tosti, Impiombato, Takahashi, Sato, Tavani, Bulgarelli, Chen, Giuliani,
  Longo, Pacciani, Argan, Barbiellini, Boffelli, Caraveo, Cattaneo, Cocco,
  Contessi, Costa, Monte, Paris, Cocco, Evangelista, Feroci, Ferrari, Fiorini,
  Froysland, Frutti, Fuschino, Galli, Gianotti, Labanti, Lapshov, Lazzarotto,
  Lipari, Marisaldi, Mastropietro, Mereghetti, Morelli, Moretti, Morselli,
  Pellizzoni, Perotti, Piano, Picozza, Pilia, Porrovecchio, Prest, Rapisarda,
  Rappoldi, Rubini, Sabatini, Scalise, Soffitta, Striani, Trifoglio, Trois,
  Vallazza, Zambra, Zanello, Pittori, Santolamazza, Verrecchia, Giommi,
  Antonelli, Colafrancesco, \& Salotti}]{Donnarumma_2009}
Donnarumma, I., Pucella, G., Vittorini, V., {et~al.} 2009, ApJ, 707, 1115

\bibitem[{Fan {et~al.}(2013)Fan, Yang, Zhang, Hua, Liu, Qin, \&
  Huang}]{Fan_2013}
Fan, J., Yang, J.~H., Zhang, J.-Y., {et~al.} 2013, PASJ, 65, 25

\bibitem[{Fan {et~al.}(2002)Fan, Cheng, \& Zhang}]{Fan_2002}
Fan, J.-H., Cheng, K.~S., \& Zhang, L. 2002, PASJ, 54, 533

\bibitem[{{Finley, C.}(2019)}]{Finley_2019}
{Finley, C.} 2019, EPJ Web Conf., 207, 02002

\bibitem[{Franckowiak {et~al.}(2020)Franckowiak, Garrappa, Paliya, Shappee,
  Stein, Strotjohann, Kowalski, Buson, Kiehlmann, Max-Moerbeck, \&
  Angioni}]{Franckowiak_2020}
Franckowiak, A., Garrappa, S., Paliya, V., {et~al.} 2020, ApJ, 893, 162

\bibitem[{Garrappa {et~al.}(2019a)Garrappa, Buson, Franckowiak, collaboration,
  Shappee, Beacom, Dong, Holoien, Kochanek, Prieto, Stanek, Thompson,
  collaboration, Aartsen, Ackermann, Adams, Aguilar, Ahlers, Ahrens, Alispach,
  Andeen, Anderson, Ansseau, Anton, Argüelles, Auffenberg, Axani, Backes,
  Bagherpour, Bai, Barbano, Barwick, Baum, Bay, Beatty, Becker, Tjus, BenZvi,
  Berley, Bernardini, Besson, Binder, Bindig, Blaufuss, Blot, Bohm, Börner,
  Böser, Botner, Bourbeau, Bourbeau, Bradascio, Braun, Bretz, Bron,
  Brostean-Kaiser, Burgman, Busse, Carver, Chen, Cheung, Chirkin, Clark,
  Classen, Collin, Conrad, Coppin, Correa, Cowen, Cross, Dave, de~André,
  Clercq, DeLaunay, Dembinski, Deoskar, Ridder, Desiati, de~Vries, de~Wasseige,
  de~With, DeYoung, Diaz, Díaz-Vélez, Dujmovic, Dunkman, Dvorak, Eberhardt,
  Ehrhardt, Eller, Evenson, Fahey, Fazely, Felde, Filimonov, Finley,
  Franckowiak, Friedman, Fritz, Gaisser, Gallagher, Ganster, Garrappa,
  Gerhardt, Ghorbani, Glauch, Glüsenkamp, Goldschmidt, Gonzalez, Grant,
  Griffith, Günder, Gündüz, Haack, Hallgren, Halve, Halzen, Hanson,
  Hebecker, Heereman, Helbing, Hellauer, Henningsen, Hickford, Hignight, Hill,
  Hoffman, Hoffmann, Hoinka, Hokanson-Fasig, Hoshina, Huang, Huber, Hultqvist,
  Hünnefeld, Hussain, In, Iovine, Ishihara, Jacobi, Japaridze, Jeong, Jero,
  Jones, Kang, Kappes, Kappesser, Karg, Karl, Karle, Katz, Kauer, Keivani,
  Kelley, Kheirandish, Kim, Kintscher, Kiryluk, Kittler, Klein, Koirala,
  Kolanoski, Köpke, Kopper, Kopper, Koskinen, Kowalski, Krings, Krückl,
  Kulacz, Kunwar, Kurahashi, Kyriacou, Labare, Lanfranchi, Larson, Lauber,
  Lazar, Leonard, Leuermann, Liu, Lohfink, Mariscal, Lu, Lucarelli, Lünemann,
  Luszczak, Madsen, Maggi, Mahn, Makino, Mallot, Mancina, Mariş, Maruyama,
  Mase, Maunu, Meagher, Medici, Medina, Meier, Meighen-Berger, Menne, Merino,
  Meures, Miarecki, Micallef, Momenté, Montaruli, Moore, Moulai, Nagai,
  Nahnhauer, Nakarmi, Naumann, Neer, Niederhausen, Nowicki, Nygren, Pollmann,
  Olivas, O’Murchadha, O’Sullivan, Palczewski, Pandya, Pankova, Park,
  Peiffer, de~los Heros, Pieloth, Pinat, Pizzuto, Plum, Price, Przybylski,
  Raab, Raissi, Rameez, Rauch, Rawlins, Rea, Reimann, Relethford, Renzi,
  Resconi, Rhode, Richman, Robertson, Rongen, Rott, Ruhe, Ryckbosch, Rysewyk,
  Safa, Herrera, Sandrock, Sandroos, Santander, Sarkar, Sarkar, Satalecka,
  Schaufel, Schlunder, Schmidt, Schneider, Schneider, Schumacher, Sclafani,
  Seckel, Seunarine, Silva, Snihur, Soedingrekso, Soldin, Song, Spiczak,
  Spiering, Stachurska, Stamatikos, Stanev, Stasik, Stein, Stettner, Steuer,
  Stezelberger, Stokstad, Stößl, Strotjohann, Stuttard, Sullivan, Sutherland,
  Taboada, Tenholt, Ter-Antonyan, Terliuk, Tilav, Tomankova, Tönnis, Toscano,
  Tosi, Tselengidou, Tung, Turcati, Turcotte, Turley, Ty, Unger, Elorrieta,
  Usner, Vandenbroucke, Driessche, van Eijk, van Eijndhoven, Vanheule, van
  Santen, Vraeghe, Walck, Wallace, Wallraff, Wandkowsky, Watson, Weaver, Weiss,
  Weldert, Wendt, Werthebach, Westerhoff, Whelan, Whitehorn, Wiebe, Wiebusch,
  Wille, Williams, Wills, Wolf, Wood, Wood, Woschnagg, Wrede, Xu, Xu, Xu,
  Yanez, Yodh, Yoshida, Yuan, \& Collaboration}]{Garrappa_2019}
Garrappa, S., Buson, S., Franckowiak, A., {et~al.} 2019a, ApJ, 880, 103

\bibitem[{Garrappa {et~al.}(2024)Garrappa, Buson, Sinapius, Franckowiak,
  Liodakis, Bartolini, Giroletti, Nanci, Principe, \& Venters}]{Garrappa_2024}
Garrappa, S., Buson, S., Sinapius, J., {et~al.} 2024, A\&A, 687, A59

\bibitem[{Giommi {et~al.}(2020)Giommi, Glauch, Padovani, Resconi, Turcati, \&
  Chang}]{Giommi_2020}
Giommi, P., Glauch, T., Padovani, P., {et~al.} 2020, MNRAS, 497, 865

\bibitem[{{Giommi, P.} {et~al.}(2020){Giommi, P.}, {Padovani, P.}, {Oikonomou,
  F.}, {Glauch, T.}, {Paiano, S.}, \& {Resconi, E.}}]{Giommi_2020_a_a}
{Giommi, P.}, {Padovani, P.}, {Oikonomou, F.}, {et~al.} 2020, A\&A, 640, L4

\bibitem[{Hassan {et~al.}(2012)Hassan, Mirabal, Contreras, \&
  Oya}]{Hassan_2012}
Hassan, T., Mirabal, N., Contreras, J.~L., \& Oya, I. 2012, MNRAS, 428, 220

\bibitem[{{H.E.S.S. Collaboration} {et~al.}(2017){H.E.S.S. Collaboration},
  {Abdalla, H.}, {Abramowski, A.}, {Aharonian, F.}, {Ait Benkhali, F.},
  {Akhperjanian, A. G.}, {Andersson, T.}, {Angüner, E. O.}, {Arrieta, M.},
  {Aubert, P.}, {Backes, M.}, {Balzer, A.}, {Barnard, M.}, {Becherini, Y.},
  {Becker Tjus, J.}, {Berge, D.}, {Bernhard, S.}, {Bernlöhr, K.}, {Blackwell,
  R.}, {Böttcher, M.}, {Boisson, C.}, {Bolmont, J.}, {Bordas, P.}, {Brun, F.},
  {Brun, P.}, {Bryan, M.}, {Bulik, T.}, {Capasso, M.}, {Carr, J.}, {Casanova,
  S.}, {Cerruti, M.}, {Chakraborty, N.}, {Chalme-Calvet, R.}, {Chaves, R. C.
  G.}, {Chen, A.}, {Chevalier, J.}, {Chrétien, M.}, {Colafrancesco, S.},
  {Cologna, G.}, {Condon, B.}, {Conrad, J.}, {Couturier, C.}, {Cui, Y.},
  {Davids, I. D.}, {Degrange, B.}, {Deil, C.}, {Devin, J.}, {deWilt, P.},
  {Dirson, L.}, {Djannati-Ataï, A.}, {Domainko, W.}, {Donath, A.}, {Drury, L.
  O’C.}, {Dubus, G.}, {Dutson, K.}, {Dyks, J.}, {Edwards, T.}, {Egberts, K.},
  {Eger, P.}, {Ernenwein, J.-P.}, {Eschbach, S.}, {Farnier, C.}, {Fegan, S.},
  {Fernandes, M. V.}, {Fiasson, A.}, {Fontaine, G.}, {Förster, A.}, {Funk,
  S.}, {Füßling, M.}, {Gabici, S.}, {Gajdus, M.}, {Gallant, Y. A.},
  {Garrigoux, T.}, {Giavitto, G.}, {Giebels, B.}, {Glicenstein, J. F.},
  {Gottschall, D.}, {Goyal, A.}, {Grondin, M.-H.}, {Hadasch, D.}, {Hahn, J.},
  {Haupt, M.}, {Hawkes, J.}, {Heinzelmann, G.}, {Henri, G.}, {Hermann, G.},
  {Hervet, O.}, {Hillert, A.}, {Hinton, J. A.}, {Hofmann, W.}, {Hoischen, C.},
  {Holler, M.}, {Horns, D.}, {Ivascenko, A.}, {Jacholkowska, A.}, {Jamrozy,
  M.}, {Janiak, M.}, {Jankowsky, D.}, {Jankowsky, F.}, {Jingo, M.}, {Jogler,
  T.}, {Jouvin, L.}, {Jung-Richardt, I.}, {Kastendieck, M. A.}, {Katarzyński,
  K.}, {Katz, U.}, {Kerszberg, D.}, {Khélifi, B.}, {Kieffer, M.}, {King, J.},
  {Klepser, S.}, {Klochkov, D.}, {Kluźniak, W.}, {Kolitzus, D.}, {Komin, Nu.},
  {Kosack, K.}, {Krakau, S.}, {Kraus, M.}, {Krayzel, F.}, {Krüger, P. P.},
  {Laffon, H.}, {Lamanna, G.}, {Lau, J.}, {Lees, J.-P.}, {Lefaucheur, J.},
  {Lefranc, V.}, {Lemière, A.}, {Lemoine-Goumard, M.}, {Lenain, J.-P.},
  {Leser, E.}, {Lohse, T.}, {Lorentz, M.}, {Liu, R.}, {López-Coto, R.},
  {Lypova, I.}, {Marandon, V.}, {Marcowith, A.}, {Mariaud, C.}, {Marx, R.},
  {Maurin, G.}, {Maxted, N.}, {Mayer, M.}, {Meintjes, P. J.}, {Meyer, M.},
  {Mitchell, A. M. W.}, {Moderski, R.}, {Mohamed, M.}, {Mohrmann, L.}, {Morå,
  K.}, {Moulin, E.}, {Murach, T.}, {de Naurois, M.}, {Niederwanger, F.},
  {Niemiec, J.}, {Oakes, L.}, {O’Brien, P.}, {Odaka, H.}, {Öttl, S.}, {Ohm,
  S.}, {Ostrowski, M.}, {Oya, I.}, {Padovani, M.}, {Panter, M.}, {Parsons, R.
  D.}, {Paz Arribas, M.}, {Pekeur, N. W.}, {Pelletier, G.}, {Perennes, C.},
  {Petrucci, P.-O.}, {Peyaud, B.}, {Pita, S.}, {Poon, H.}, {Prokhorov, D.},
  {Prokoph, H.}, {Pühlhofer, G.}, {Punch, M.}, {Quirrenbach, A.}, {Raab, S.},
  {Reimer, A.}, {Reimer, O.}, {Renaud, M.}, {de los Reyes, R.}, {Rieger, F.},
  {Romoli, C.}, {Rosier-Lees, S.}, {Rowell, G.}, {Rudak, B.}, {Rulten, C. B.},
  {Sahakian, V.}, {Salek, D.}, {Sanchez, D. A.}, {Santangelo, A.}, {Sasaki,
  M.}, {Schlickeiser, R.}, {Schüssler, F.}, {Schulz, A.}, {Schwanke, U.},
  {Schwemmer, S.}, {Settimo, M.}, {Seyffert, A. S.}, {Shafi, N.}, {Shilon, I.},
  {Simoni, R.}, {Sol, H.}, {Spanier, F.}, {Spengler, G.}, {Spies, F.},
  {Stawarz, Ł.}, {Steenkamp, R.}, {Stegmann, C.}, {Stinzing, F.}, {Stycz, K.},
  {Sushch, I.}, {Tavernet, J.-P.}, {Tavernier, T.}, {Taylor, A. M.}, {Terrier,
  R.}, {Tibaldo, L.}, {Tiziani, D.}, {Tluczykont, M.}, {Trichard, C.}, {Tuffs,
  R.}, {Uchiyama, Y.}, {van der Walt, D. J.}, {van Eldik, C.}, {van Soelen,
  B.}, {Vasileiadis, G.}, {Veh, J.}, {Venter, C.}, {Viana, A.}, {Vincent, P.},
  {Vink, J.}, {Voisin, F.}, {Völk, H. J.}, {Vuillaume, T.}, {Wadiasingh, Z.},
  {Wagner, S. J.}, {Wagner, P.}, {Wagner, R. M.}, {White, R.}, {Wierzcholska,
  A.}, {Willmann, P.}, {Wörnlein, A.}, {Wouters, D.}, {Yang, R.}, {Zabalza,
  V.}, {Zaborov, D.}, {Zacharias, M.}, {Zdziarski, A. A.}, {Zech, A.}, {Zefi,
  F.}, {Ziegler, A.}, {Żywucka, N.}, {LAT Collaboration}, {Ackermann, M.},
  {Ajello, M.}, {Baldini, L.}, {Barbiellini, G.}, {Bellazzini, R.}, {Blandford,
  R. D.}, {Bonino, R.}, {Bregeon, J.}, {Bruel, P.}, {Buehler, R.}, {Caliandro,
  G. A.}, {Cameron, R. A.}, {Caragiulo, M.}, {Caraveo, P. A.}, {Cavazzuti, E.},
  {Cecchi, C.}, {Chiang, J.}, {Chiaro, G.}, {Ciprini, S.}, {Cohen-Tanugi, J.},
  {Costanza, F.}, {Cutini, S.}, {D’Ammando, F.}, {de Palma, F.}, {Desiante,
  R.}, {Di Lalla, N.}, {Di Mauro, M.}, {Di Venere, L.}, {Donaggio, B.},
  {Favuzzi, C.}, {Focke, W. B.}, {Fusco, P.}, {Gargano, F.}, {Gasparrini, D.},
  {Giglietto, N.}, {Giordano, F.}, {Giroletti, M.}, {Guillemot, L.}, {Guiriec,
  S.}, {Horan, D.}, {Jóhannesson, G.}, {Kamae, T.}, {Kensei, S.}, {Kocevski,
  D.}, {Larsson, S.}, {Li, J.}, {Longo, F.}, {Loparco, F.}, {Lovellette, M.
  N.}, {Lubrano, P.}, {Maldera, S.}, {Manfreda, A.}, {Mazziotta, M. N.},
  {Michelson, P. F.}, {Mizuno, T.}, {Monzani, M. E.}, {Morselli, A.}, {Negro,
  M.}, {Nuss, E.}, {Orienti, M.}, {Orlando, E.}, {Paneque, D.}, {Perkins, J.
  S.}, {Pesce-Rollins, M.}, {Piron, F.}, {Pivato, G.}, {Porter, T. A.},
  {Principe, G.}, {Rainò, S.}, {Razzano, M.}, {Simone, D.}, {Siskind, E. J.},
  {Spada, F.}, {Spinelli, P.}, {Thayer, J. B.}, {Torres, D. F.}, {Torresi, E.},
  {Troja, E.}, {Vianello, G.}, \& {Wood, K. S.}}]{Abdalla_2017}
{H.E.S.S. Collaboration}, {Abdalla, H.}, {Abramowski, A.}, {et~al.} 2017, A\&A,
  600, A89

\bibitem[{Itoh {et~al.}(2016)Itoh, Nalewajko, Fukazawa, Uemura, Tanaka,
  Kawabata, Madejski, Schinzel, Kanda, Shiki, Akitaya, Kawabata, Moritani,
  Nakaoka, Ohsugi, Sasada, Takaki, Takata, Ui, Yamanaka, \&
  Yoshida}]{Itoh_2016}
Itoh, R., Nalewajko, K., Fukazawa, Y., {et~al.} 2016, ApJ, 833, 77

\bibitem[{Kadler {et~al.}(2016)}]{Kadler_2016}
Kadler, M. {et~al.} 2016, Nature Phys., 12, 807

\bibitem[{Kotera \& Olinto(2011)}]{Kotera_2011}
Kotera, K. \& Olinto, A.~V. 2011, ARA\&A, 49, 119–153

\bibitem[{{Krauß, F.} {et~al.}(2018){Krauß, F.}, {Deoskar, K.}, {Baxter, C.},
  {Kadler, M.}, {Kreter, M.}, {Langejahn, M.}, {Mannheim, K.}, {Polko, P.},
  {Wang, B.}, \& {Wilms, J.}}]{Krauss_2018}
{Krauß, F.}, {Deoskar, K.}, {Baxter, C.}, {et~al.} 2018, A\&A, 620, A174

\bibitem[{Liao {et~al.}(2022)Liao, Sheng, Jiang, Chang, Wang, Xu, Shu, Fan, \&
  Wang}]{Liao_2022}
Liao, N.-H., Sheng, Z.-F., Jiang, N., {et~al.} 2022, ApJL, 932, L25

\bibitem[{{Mannheim}(1993)}]{Mannheim_1993}
{Mannheim}, K. 1993, A\&A, 269, 67

\bibitem[{Massaro {et~al.}(2006)Massaro, Tramacere, Perri, Giommi, \&
  Tosti}]{Massaro_2006}
Massaro, E., Tramacere, A., Perri, M., Giommi, P., \& Tosti, G. 2006, A\&A,
  448, 861

\bibitem[{Mattox {et~al.}(1996)Mattox, Bertsch, Chiang, Dingus, Digel,
  Esposito, Fierro, Hartman, Hunter, Kanbach, Kniffen, Lin, Macomb,
  Mayer-Hasselwander, Michelson, von Montigny, Mukherjee, Nolan, Ramanamurthy,
  Schneid, Sreekumar, Thompson, \& Willis}]{Mattox_1996}
Mattox, J.~R., Bertsch, D.~L., Chiang, J., {et~al.} 1996, ApJ, 461, 396

\bibitem[{Murase {et~al.}(2013)Murase, Ahlers, \& Lacki}]{Murase@2013}
Murase, K., Ahlers, M., \& Lacki, B.~C. 2013, Phys. Rev. D, 88, 121301

\bibitem[{Murase {et~al.}(2012)Murase, Dermer, Takami, \&
  Migliori}]{Murase_2012}
Murase, K., Dermer, C.~D., Takami, H., \& Migliori, G. 2012, ApJ, 749, 63

\bibitem[{Murase \& Stecker(2023)}]{Murase_2023}
Murase, K. \& Stecker, F.~W. 2023, High-Energy Neutrinos from Active Galactic
  Nuclei (WORLD SCIENTIFIC), 483–540

\bibitem[{Mücke {et~al.}(2003)Mücke, Protheroe, Engel, Rachen, \&
  Stanev}]{MUCKE_2003}
Mücke, A., Protheroe, R., Engel, R., Rachen, J., \& Stanev, T. 2003,
  Astroparticle Physics, 18, 593

\bibitem[{Nalewajko(2013)}]{Nalewajko_2013}
Nalewajko, K. 2013, MNRAS, 430, 1324

\bibitem[{Noel {et~al.}(2022)Noel, Gaur, Gupta, Wierzcholska, Ostrowski,
  Dhiman, \& Bhatta}]{Noel_2022}
Noel, A.~P., Gaur, H., Gupta, A.~C., {et~al.} 2022, ApJS, 262, 4

\bibitem[{Padovani {et~al.}(2018)Padovani, Giommi, Resconi, Glauch, Arsioli,
  Sahakyan, \& Huber}]{Padovani_2018}
Padovani, P., Giommi, P., Resconi, E., {et~al.} 2018, MNRAS, 480, 192

\bibitem[{Prince {et~al.}(2018)Prince, Raman, Hahn, Gupta, \&
  Majumdar}]{Prince_2018}
Prince, R., Raman, G., Hahn, J., Gupta, N., \& Majumdar, P. 2018, ApJ, 866, 16

\bibitem[{Rajput {et~al.}(2020)Rajput, Stalin, \& Suvendu}]{Rajput_2020}
Rajput, B., Stalin, C.~S., \& Suvendu, R. 2020, A\&A, 634, A80

\bibitem[{Rani {et~al.}(2016)Rani, Stalin, \& Rakshit}]{Rani_2016}
Rani, P., Stalin, C.~S., \& Rakshit, S. 2016, MNRAS, 466, 3309

\bibitem[{Reines(1960)}]{Reines_1960}
Reines, F. 1960, Annual Review of Nuclear Science, 10, 1

\bibitem[{Rodrigues {et~al.}(2018)Rodrigues, Fedynitch, Gao, Boncioli, \&
  Winter}]{Rodrigues_2018}
Rodrigues, X., Fedynitch, A., Gao, S., Boncioli, D., \& Winter, W. 2018, ApJ,
  854, 54

\bibitem[{Stecker {et~al.}(1991)Stecker, Done, Salamon, \&
  Sommers}]{Stecker_1991}
Stecker, F.~W., Done, C., Salamon, M.~H., \& Sommers, P. 1991, Phys. Rev.
  Lett., 66, 2697

\bibitem[{Urry \& Padovani(1995)}]{Urry_1995}
Urry, C.~M. \& Padovani, P. 1995, PASP, 107, 803

\bibitem[{Vaughan {et~al.}(2003)Vaughan, Edelson, Warwick, \&
  Uttley}]{Vaughan_2003}
Vaughan, S., Edelson, R., Warwick, R.~S., \& Uttley, P. 2003, MNRAS, 345, 1271

\bibitem[{Vercellone {et~al.}(2010)Vercellone, D'Ammando, Vittorini,
  Donnarumma, Pucella, Tavani, Ferrari, Raiteri, Villata, Romano, Krimm,
  Tiengo, Chen, Giovannini, Venturi, Giroletti, Kovalev, Sokolovsky, Pushkarev,
  Lister, Argan, Barbiellini, Bulgarelli, Caraveo, Cattaneo, Cocco, Costa,
  Monte, Paris, Cocco, Evangelista, Feroci, Fiorini, Fornari, Froysland,
  Fuschino, Galli, Gianotti, Labanti, Lapshov, Lazzarotto, Lipari, Longo,
  Giuliani, Marisaldi, Mereghetti, Morselli, Pellizzoni, Pacciani, Perotti,
  Piano, Picozza, Pilia, Prest, Rapisarda, Rappoldi, Sabatini, Soffitta,
  Striani, Trifoglio, Trois, Vallazza, Zambra, Zanello, Pittori, Verrecchia,
  Santolamazza, Giommi, Colafrancesco, Salotti, Agudo, Aller, Aller, Arkharov,
  Bach, Bachev, Beltrame, Benítez, Böttcher, Buemi, Calcidese, Capezzali,
  Carosati, Chen, Rio, Paola, Dolci, Dultzin, Forné, Gómez, Gurwell,
  Hagen-Thorn, Halkola, Heidt, Hiriart, Hovatta, Hsiao, Jorstad, Kimeridze,
  Konstantinova, Kopatskaya, Koptelova, Kurtanidze, Lähteenmäki, Larionov,
  Leto, Ligustri, Lindfors, Lopez, Marscher, Mujica, Nikolashvili, Nilsson,
  Mommert, Palma, Pasanen, Roca-Sogorb, Ros, Roustazadeh, Sadun, Saino, Sigua,
  Sorcia, Takalo, Tornikoski, Trigilio, Turchetti, \& Umana}]{Vercellone_2010}
Vercellone, S., D'Ammando, F., Vittorini, V., {et~al.} 2010, ApJ, 712, 405

\bibitem[{Winter(2013)}]{Winter_2013}
Winter, W. 2013, Phys. Rev. D, 88, 083007

\bibitem[{Xiao {et~al.}(2024)Xiao, Yang, Zhang, Zhang, Fan, Fu, \&
  Yang}]{Xiao_2024}
Xiao, H., Yang, W., Zhang, Y., {et~al.} 2024, ApJ, 966, 99

\end{thebibliography}
\clearpage 
\onecolumn
\appendix 
\renewcommand{\thefigure}{A.\arabic{figure}}
\setcounter{figure}{0}
\section*{{Appendix A: Histogram for total coincident sources from uniform RA and Dec distributions and histogram of randomly generated declination values}}

\begin{figure}[!htbp]
  \centering
  \begin{minipage}{0.48\textwidth}
    \centering
    \includegraphics[width=\textwidth]{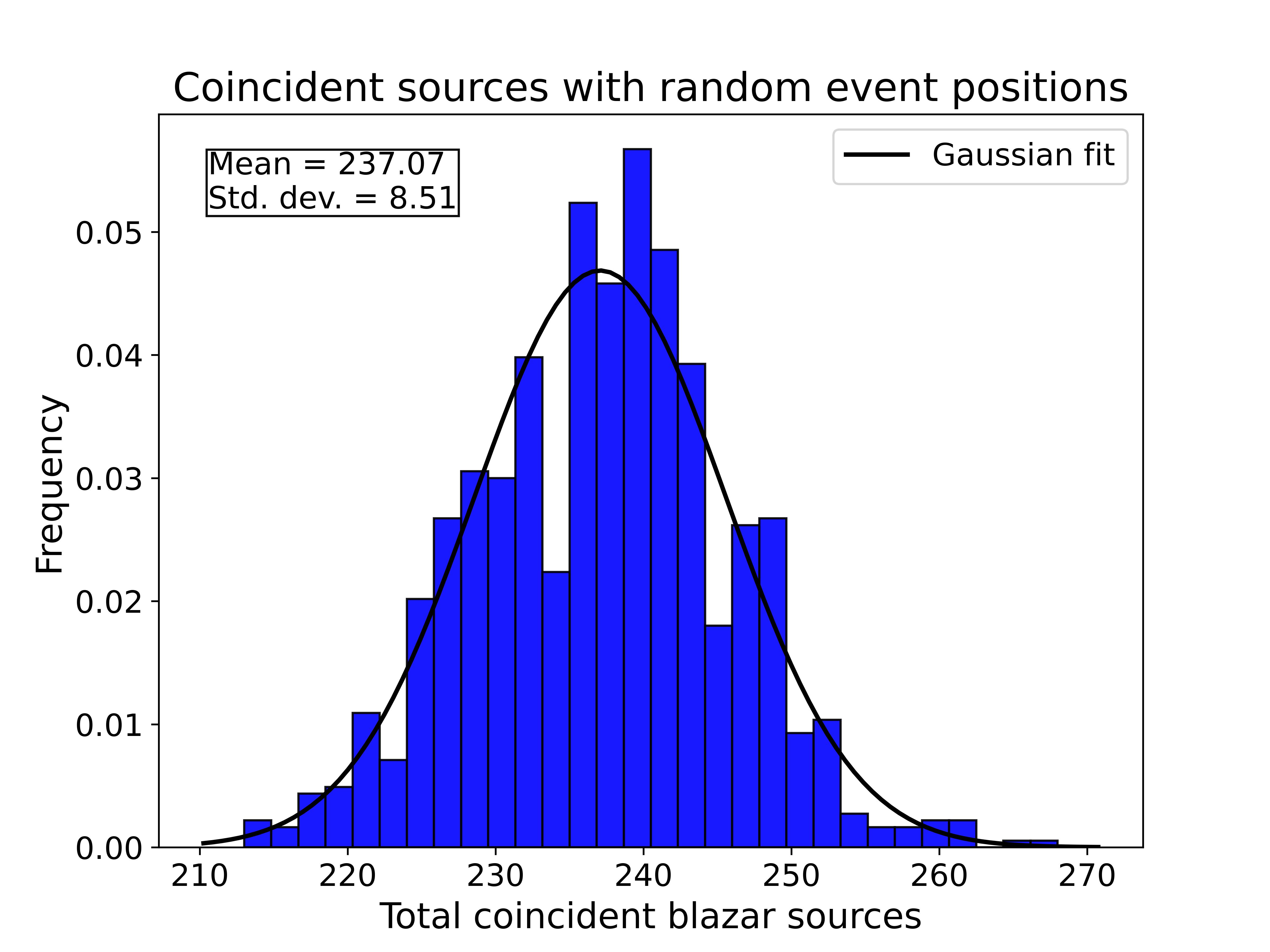}
  \end{minipage}
  \hfill
  \begin{minipage}{0.48\textwidth}
    \centering
    \includegraphics[width=\textwidth]{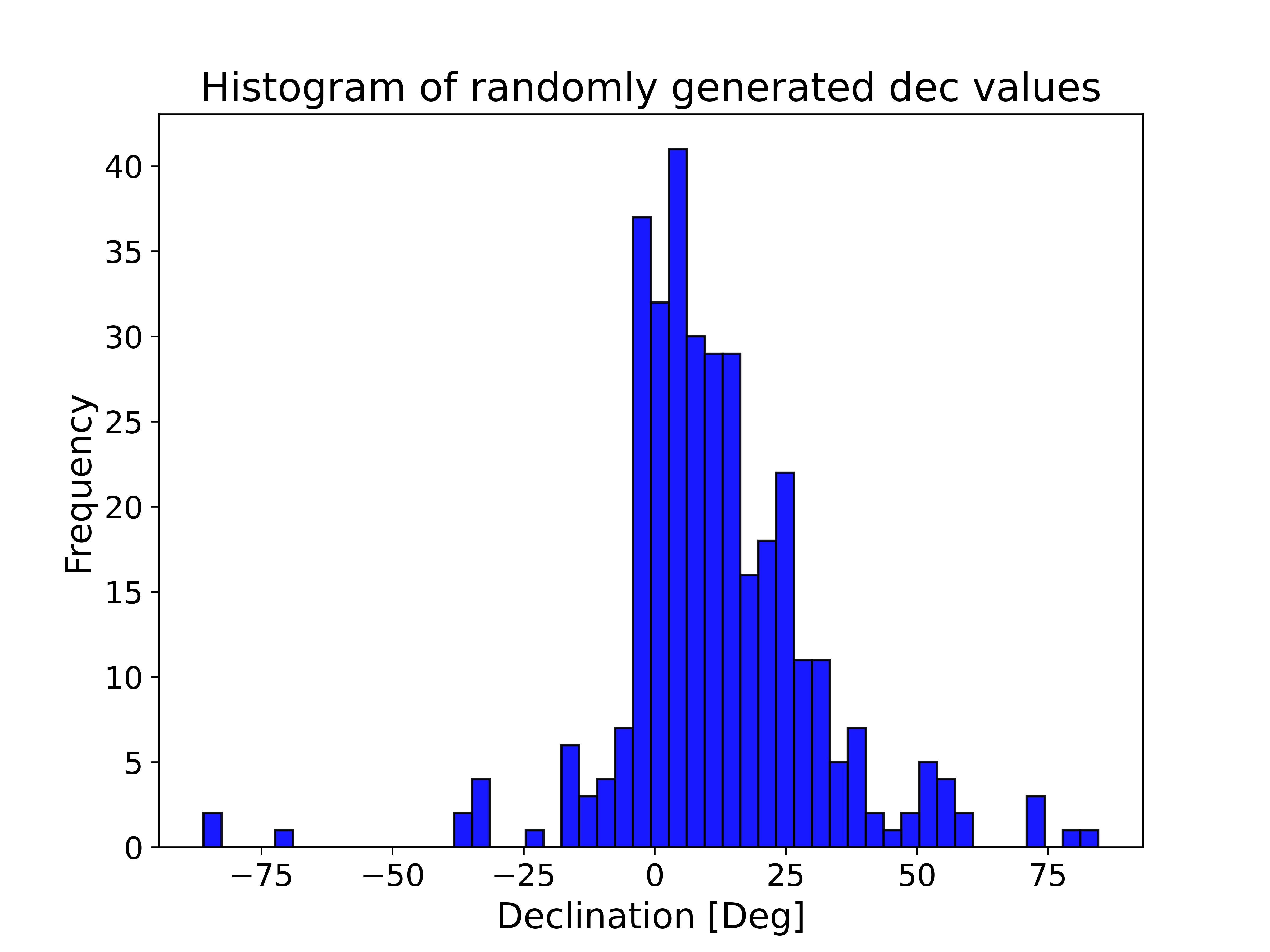}
  \end{minipage}
  
  \caption{Left panel: Histogram of total coincident blazar sources with the event positions generated from random uniform RA and Dec.  
  Right panel: Histogram of randomly generated non-uniform declination values, evaluated from the declination values of real astrophysical neutrino events.}
  \label{figA1A2}
\end{figure}

\clearpage

\begin{appendix}

\renewcommand{\thetable}{B.\arabic{table}}
\setcounter{table}{0}
\section*{{Appendix B: Spectral parameters for different sources in different time episodes}}\label{appendix:B}

\begin{table}[htbp]
    \centering
    \caption{Time duration, photon flux and index of the source 4FGL J0506.9+0323.  The spectrum type in each observational episode is Powerlaw.}
    \begin{tabular}{cccc}
    \hline
   \textbf{Time duration} & \textbf{Photon flux}  & \textbf{Index}   \\
     {\textbf{[MJD]}} & \textbf{($\times 10^{-9}$ cm$^{-2}$ s$^{-1}$)}   \\
    \hline
   54700--55600 & $(1.1\pm0.73)$ &  $-1.69\pm0.21$ \\
    55600--56500 & $(2.67\pm1.42)$ & $-1.87\pm0.19$ \\
    56500--57400 & $(4.62\pm2.42)$ &  $-2.1\pm0.21$ \\
    57400--58300 & $ (2.64\pm2.49)$ & $-2.28\pm0.37$ \\
    58300--59200 & $(2.99\pm1.85)$  &  $-2.04\pm0.23$ \\
    59200--60100 & $(5.04\pm2.53) $  &  $-2.32\pm0.27$ \\
    60100--60370 & $(3.48\pm4.12)$ & $-2.10\pm0.47$ \\
    \hline
    \end{tabular}
\end{table}

\begin{table}[htbp]
    \centering
    \caption{Time duration, photon flux and index of the source 4FGL J1012.3+0629.  The spectrum type in each observational episode is Powerlaw.  }
    \begin{tabular}{cccc}
     \hline
     \textbf{Time duration} & \textbf{Photon flux}  & \textbf{Index}   \\
     {\textbf{[MJD]}} & \textbf{($\times 10^{-9}$ cm$^{-2}$ s$^{-1}$)} \\
     \hline
   54700--55600 & $(8.30\pm1.94)$   & $-2.12\pm0.11$ \\
    55600--56500 & $(11.13\pm2.16)$   & $-2.11\pm0.09$ \\
    56500--57400 & $(14.23\pm2.14)$   &  $-2.1\pm0.08$ \\
    57400--58300 & $ (13.08\pm2.00)$  & $-2.07\pm0.07$ \\
    58300--59200 & $(17.04\pm2.05)$  & $-2.00\pm0.06$ \\
    59200--60100 & $(11.91\pm1.96) $ &  $-2.06\pm0.08$ \\
    60100--60370 & $(12.28\pm3.98)$ &  $-2.3\pm0.19$ \\
    \hline
    \end{tabular}
\end{table}

\begin{table}[htbp]
    \centering
    \caption{Time duration, photon flux and index of the source 4FGL J1016.0+0512.  The spectrum type in each observational episode is Powerlaw.}
    \begin{tabular}{cccc}
    \hline
  \textbf{Time duration} & \textbf{Photon flux}  & \textbf{Index}   \\
     {\textbf{[MJD]}} & \textbf{($\times 10^{-9}$ cm$^{-2}$ s$^{-1}$)} \\
   \hline
   54700--55600 & $(45.95\pm2.56)$ & $-2.12\pm0.03$ \\
    55600--56500 & $(6.86\pm2.65)$ & $ -2.25\pm0.18$ \\
    56500--57400 & $(1.54\pm1.21)$ & $ -1.90\pm0.28$ \\
    57400--58300 & $ (10.3\pm2.62)$ & $ -2.6\pm0.16$ \\
    58300--59200 & $(3.11\pm2.03)$ & $ -2.24\pm0.27$ \\
    59200--60100 & $(3.38\pm2.91) $ & $ -2.47\pm0.39$\\
    60100--60370 & $(0.0003\pm0.01)$ & $-2.21\pm21.41$ \\
     \hline
    \end{tabular}
\end{table}

\begin{table}[htbp]
    \centering
    \caption{Time duration, photon flux and index of the source 4FGL J1018.4+0528.  The spectrum type in each observational episode is Powerlaw.}
    \begin{tabular}{cccc}
    \hline
   \textbf{Time duration} & \textbf{Photon flux}  & \textbf{Index}   \\
     {\textbf{[MJD]}} & \textbf{($\times 10^{-9}$ cm$^{-2}$ s$^{-1}$)} \\
     \hline
   54700--55600 & $(12.56\pm2.29)$ &   $-2.22\pm0.09$ \\
    55600--56500 & $(2.81\pm1.99)$  & $-2.15\pm0.29$\\
    56500--57400 & $(0.25\pm0.51)$ & $-1.69\pm0.60$ \\
    57400--58300 & $ (0.15\pm0.20)$  & $ -1.29\pm0.45$ \\
    58300--59200 & $(11.06\pm2.89)$  & $-2.5\pm0.17$ \\
    59200--60100 & $(3.49\pm2.10) $  & $ -2.33\pm0.28$ \\
    60100--60370 & $(5.08\pm4.07)$  & $ -2.35\pm0.40$ \\
   \hline
    \end{tabular}
\end{table}

\begin{table}[htbp]
    \centering
    \caption{Time duration, photon flux and index of the source 4FGL J2118.0+0019.  The spectrum type in each observational episode is Powerlaw.}
    \begin{tabular}{cccc}
    \hline
   \textbf{Time duration} & \textbf{Photon flux}  & \textbf{Index}   \\
     {\textbf{[MJD]}} & \textbf{($\times 10^{-9}$ cm$^{-2}$ s$^{-1}$)} \\
    \hline
   54700--55600 & $(9.98\pm2.41)$  & $-2.41\pm0.13$ \\
    55600--56500 & $(2.55\pm1.95)$ & $-2.14\pm0.28$ \\
    56500--57400 & $(0.98\pm1.26)$  & $-2.01\pm0.45$ \\
    57400--58300 & $ (2.75\pm1.63)$  & $-2.12\pm0.22$ \\
    58300--59200 & $(6.12\pm2.56)$   & $-2.44\pm0.22$ \\
    59200--60100 & $(3.92\pm2.41) $  & $ -2.26\pm0.27$ \\
    60100--60370 & $(6.13\pm3.21)$ & $-2.09\pm0.23$ \\
    \hline
    \end{tabular}
\end{table}

\begin{table}[htbp]
    \centering
    \caption{Time duration, photon flux and index of the source  4FGL J2223.3+0102.  The spectrum type in each observational episode is Powerlaw.}
    \begin{tabular}{cccc}
    \hline
  \textbf{Time duration} & \textbf{Photon flux}  & \textbf{Index}   \\
     {\textbf{[MJD]}} & \textbf{($\times 10^{-9}$ cm$^{-2}$ s$^{-1}$)} \\
    \hline
   54700--55600 & $(1.74\pm1.11)$  & $ -1.77\pm0.23$ \\
    55600--56500 & $(2.53\pm2.16)$  & $-2.39\pm0.65$ \\
    56500--57400 & $(0.51\pm0.60)$  & $ -1.69\pm0.35$ \\
    57400--58300 & $ (0.80\pm0.85)$  & $ -1.81\pm0.35$ \\
    58300--59200 & $(0.39\pm0.18)$   & $ -1.89\pm0.15$ \\
    59200--60100 & $(8.06\pm12.80) $  & $-2.68\pm1.93$ \\
    60100--60370 & $(14.02\pm0.67)$  & $ -3.28\pm0.01$ \\
     \hline
    \end{tabular}
\end{table}

\begin{table}[htbp]
    \centering
    \caption{Time duration, photon flux, spectral index and curvature parameter of the source  4FGL J2226.8+0051. The spectrum type in each observational episode is Logparabola.}
    \begin{tabular}{ccccc}
    \hline
    \textbf{Time duration} & \textbf{Photon flux}  & \textbf{$\alpha$} & \textbf{$\beta$}    \\
      {\textbf{[MJD]}} & \textbf{($\times 10^{-9}$ cm$^{-2}$ s$^{-1}$)}   \\
    \hline
   54700--55600 & $(3.38\pm2.53)$ & $2.46\pm0.35$ &  $0.24\pm0.23$ \\
    55600--56500 & $(2.96\pm4.10)$ &  $3.18\pm4.75$ & $1.00\pm3.52$  \\
    56500--57400 & $(1.75\pm1.81)$  & $ 2.34\pm0.54$ & $0.43\pm     0.40$ \\
    57400--58300 & $ (4.37\pm2.11)$  & $4.2\pm3.7$ & $ 1.07\pm1.38$\\
    58300--59200 & $(2.92\pm1.83)$  & $ 3.16\pm0.95$ & $ 0.77\pm    0.52$ \\
    59200--60100 & $(6.03\pm2.59) $ & $ 2.31\pm0.22$ & $ 0.27\pm      0.17$ \\
    60100--60370 & $(0.91\pm0.84)$ & $ 1.5\pm0.78$ & $0.59\pm0.64$ \\
     \hline
    \end{tabular}
\end{table}

\begin{table}[htbp]
    \centering
    \caption{Time duration, photon flux and index of the source   4FGL J2227.9+0036. The spectrum type in each observational episode is Powerlaw.}
    \begin{tabular}{ccccc}
    \hline
   \textbf{Time duration} & \textbf{Photon flux}  & \textbf{Index}   \\
     {\textbf{[MJD]}} & \textbf{($\times 10^{-9}$ cm$^{-2}$ s$^{-1}$)} \\
    \hline
   54700--55600 & $(4.55\pm1.90)$ & $-2.19\pm0.21$  \\
    55600--56500 & $(9.17\pm1.80)$  & $ -1.99\pm0.09$  \\
    56500--57400 & $(5.13\pm1.67)$  & $ -1.91\pm0.13$ \\
    57400--58300 & $ (8.70\pm1.79)$ & $-1.91\pm0.09$ \\
    58300--59200 & $(9.92\pm1.80)$  & $-1.93\pm0.08$  \\
    59200--60100 & $(2.54\pm0.96) $ & $ -1.74\pm0.14$ \\
    60100--60370 & $(1.81\pm1.81)$   & $-1.77\pm0.33$ \\
     \hline
    \end{tabular}
\end{table}

\begin{table}[htbp]
    \centering
    \caption{Time duration, photon flux and index of the source   4FGL J2252.6+1245. The spectrum type in each observational episode is Powerlaw.}
    \begin{tabular}{cccc}
    \hline
   \textbf{Time duration} & \textbf{Photon flux}  & \textbf{Index}   \\
     {\textbf{[MJD]}} & \textbf{($\times 10^{-9}$ cm$^{-2}$ s$^{-1}$)} \\
    \hline
   54700--55600 & $(3.62\pm2.36)$   & $-2.40\pm0.27$\\
    55600--56500 & $(0.11\pm0.17)$  & $-1.29\pm0.46$ \\
    56500--57400 & $(1.57\pm1.03)$  & $ -1.8\pm0.22$ \\
    57400--58300 & $ (1.12\pm1.28)$  & $-1.9\pm0.40$ \\
    58300--59200 & $(0.32\pm0.48)$   & $ -1.60\pm0.43$ \\
    59200--60100 & $(1.02\pm0.95) $  & $ -1.88\pm0.32$ \\
    60100--60370 & $(3.14\pm3.11)$  & $-2.19\pm0.42$ \\
    \hline
    \end{tabular}
\end{table}
\end{appendix}

 \clearpage

\begin{appendix}

\renewcommand{\thefigure}{C.\arabic{figure}}
\setcounter{figure}{0}

\section*{{Appendix C: Light curves for sources which are unlikely to be high-energy neutrino emitters}} \label{appendix:C}

\begin{figure}[!htbp]
  \centering
  
   \includegraphics[width=0.46\textwidth]{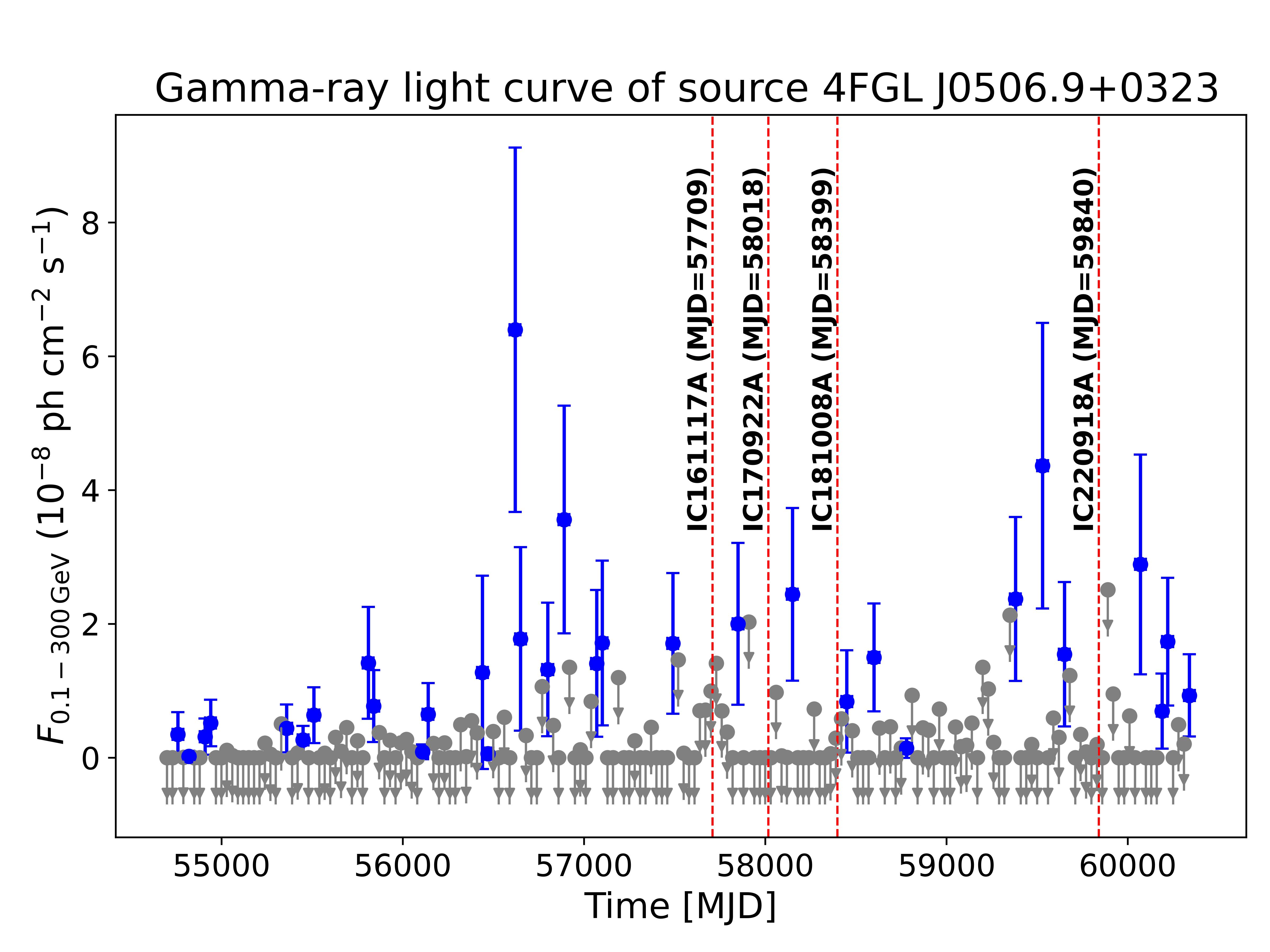}
  \includegraphics[width=0.46\textwidth]{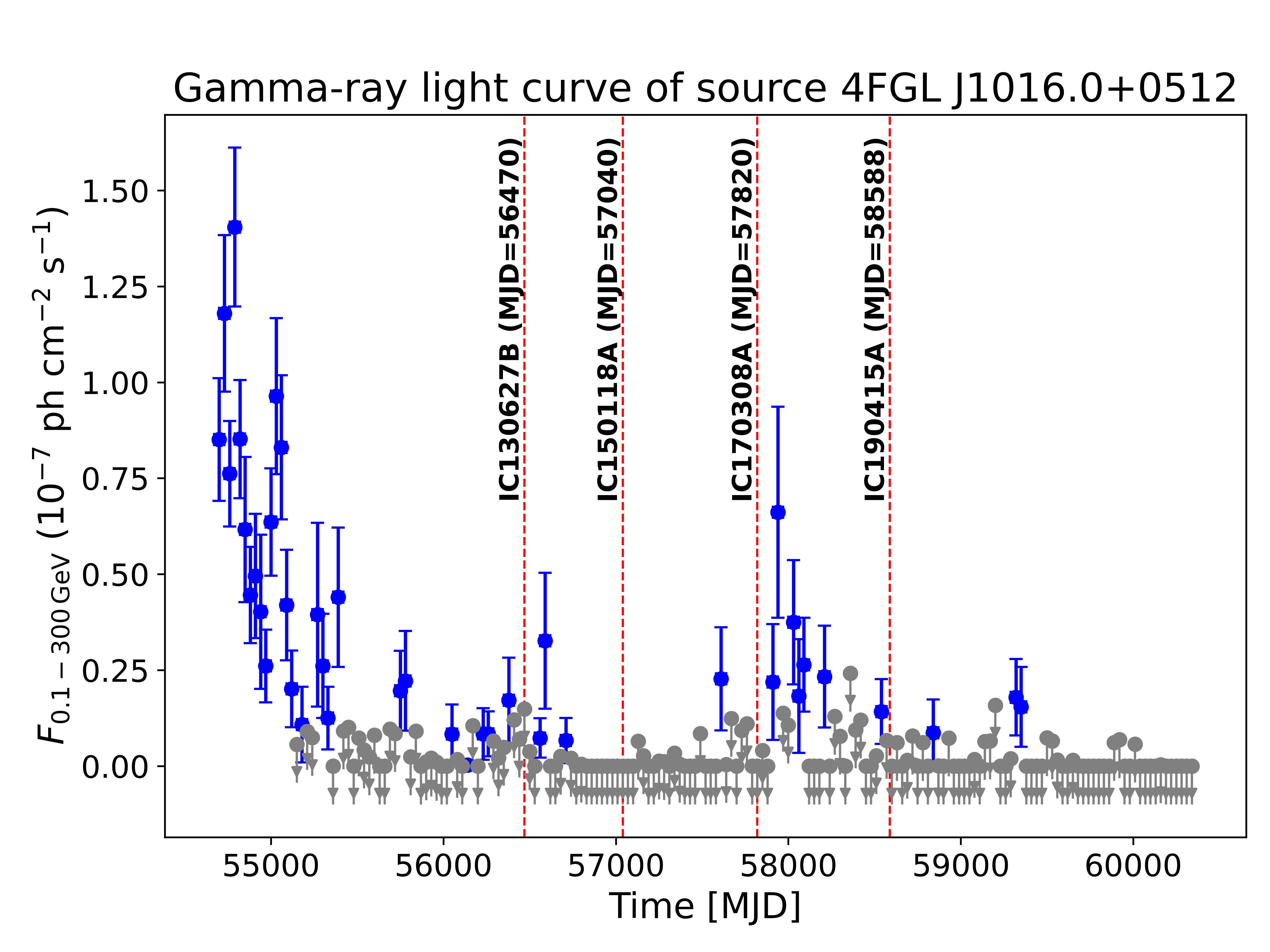}
  
  \caption{$\gamma$-ray light curve of the sources 4FGL J0506.9+0323 and  4FGL J1016.0+0512  from  54700 to 60370 MJD  in 30-days time bins.}
\end{figure}

\begin{figure}[!htbp]
  \centering
  
   \includegraphics[width=0.46\textwidth]{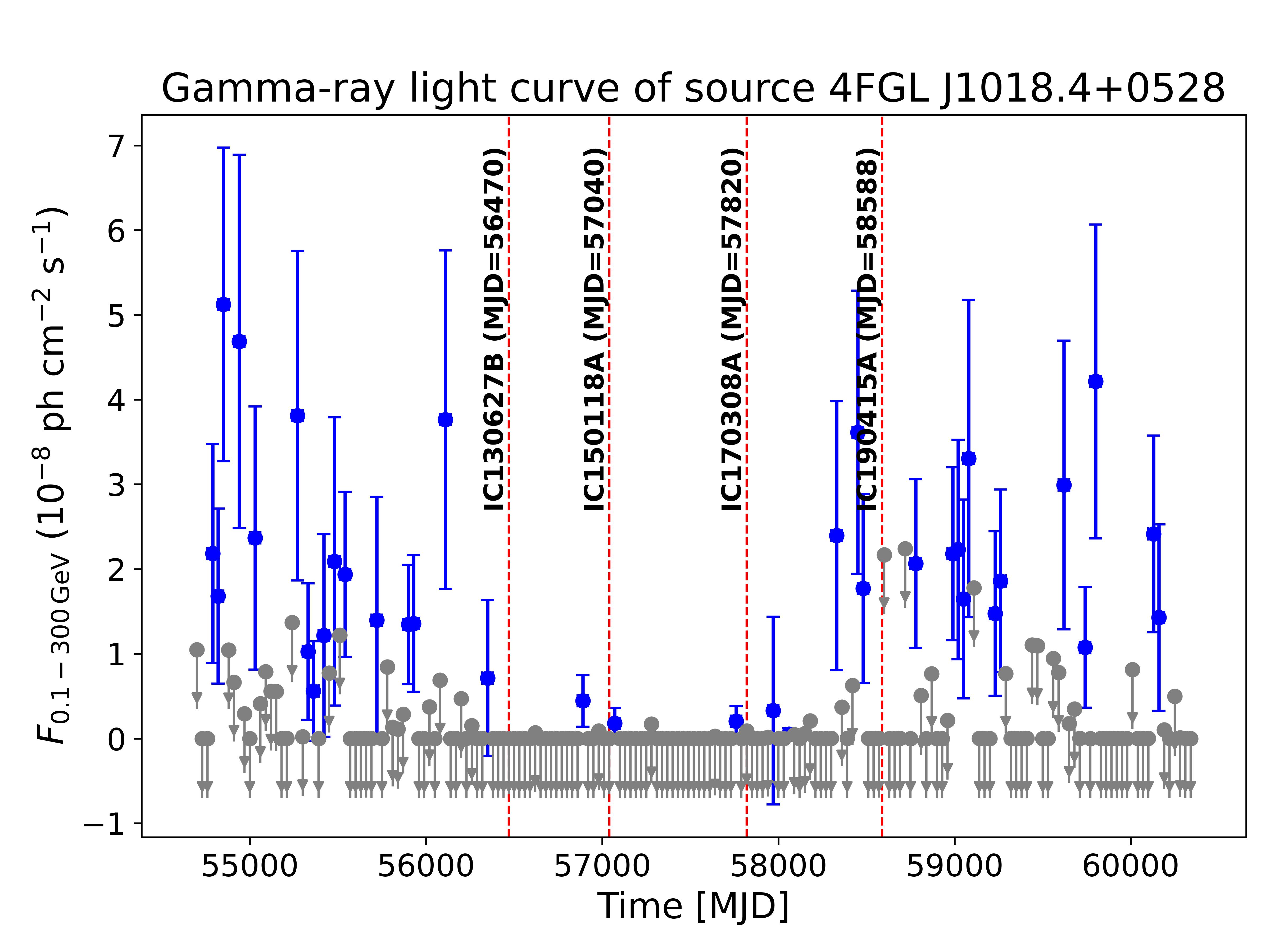}
  \includegraphics[width=0.46\textwidth]{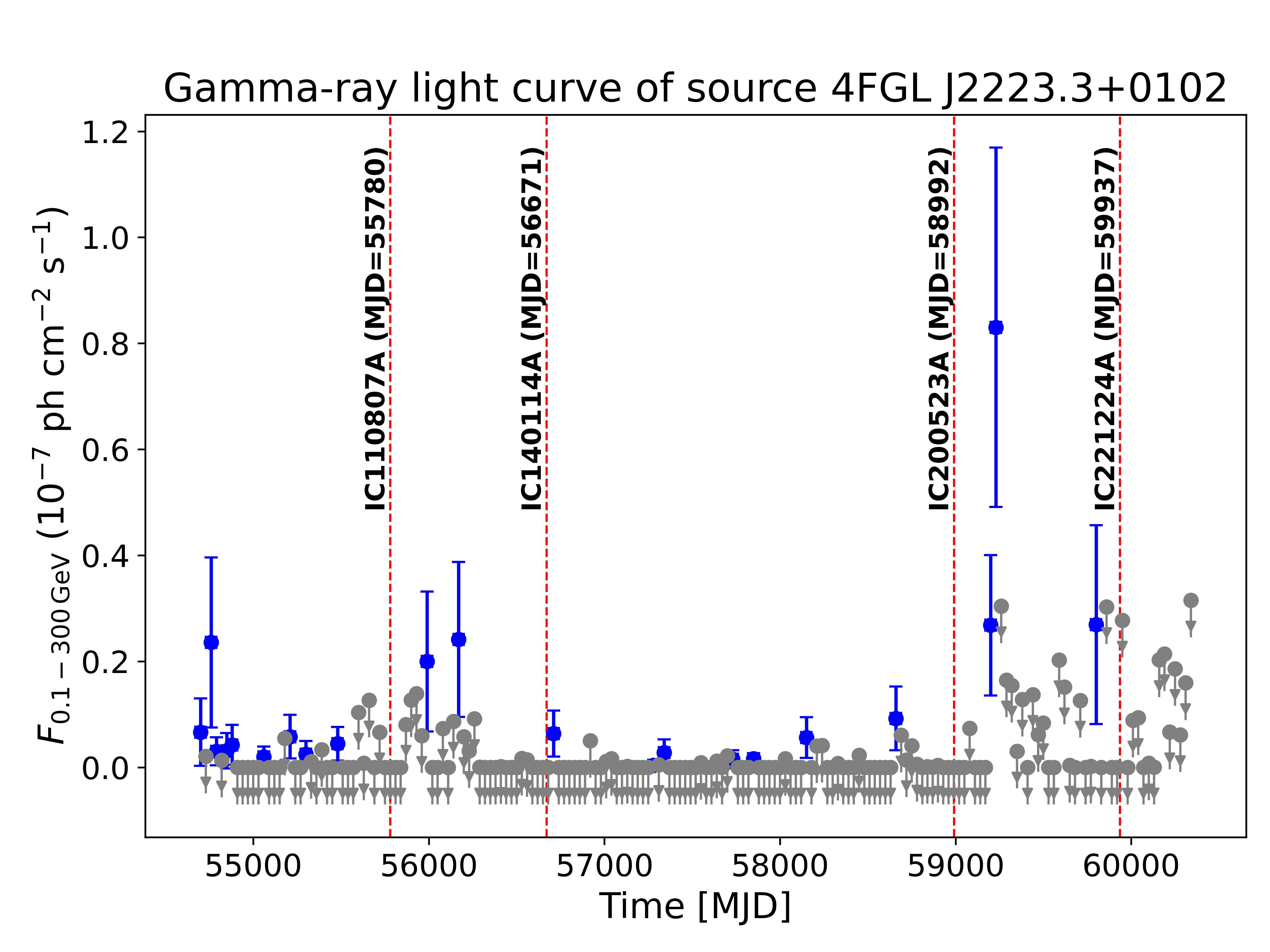}
  
  \caption{$\gamma$-ray light curve of the sources 4FGL J1018.4+0528 and 4FGL J2223.3+0102  from  54700 to 60370 MJD  in 30-days time bins.}
\end{figure}

\begin{figure}[!htbp]
  \centering
  
   \includegraphics[width=0.46\textwidth]{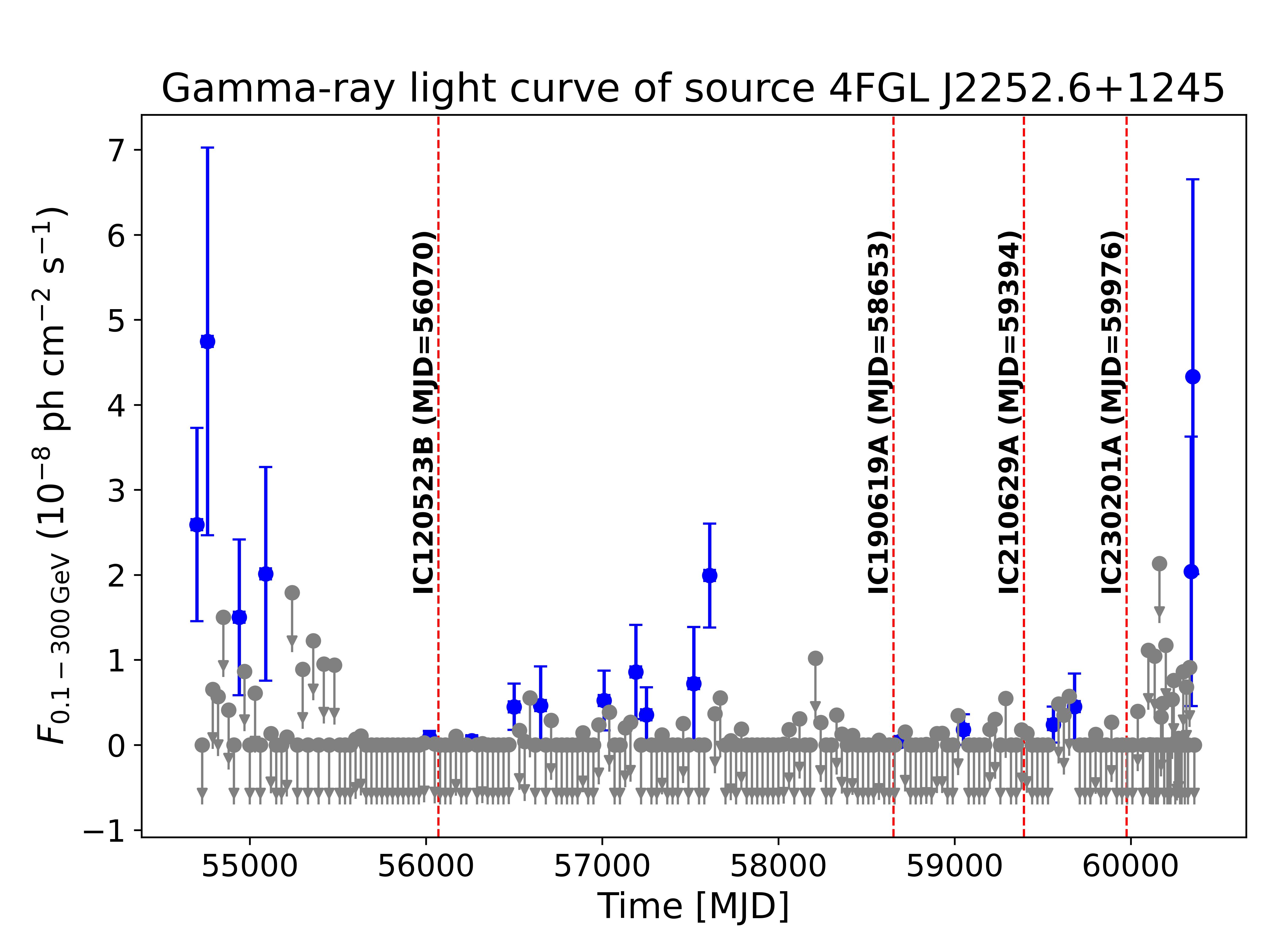}

  \caption{$\gamma$-ray light curve of the source 4FGL J2252.6+1245 from  54700 to 60370 MJD  in 30-days time bins.}
\end{figure}
\end{appendix}

\clearpage

\begin{appendix}

\renewcommand{\thetable}{D.\arabic{table}}
\setcounter{table}{0}
\section*{Appendix D: Photon flux and variability index of nine blazar sources.} \label{appendix:D}
\begin{table}[htbp]
\centering
   \caption[]{Photon flux and variability index of nine blazar sources, each containing four neutrino events. These values are obtained from the 4FGL-DR4 catalog.}
   \label{tab:your_table_label} 
   
   \begin{tabular}{lcc}
      \hline
      \noalign{\smallskip}
      \textbf{Source (RA$^{\circ}$, Dec$^{\circ}$)} & \textbf{Photon flux(cm$^{-2}\ $s$^{-1}$)} & \textbf{Variability index}  \\
      \noalign{\smallskip}
      \hline
      \noalign{\smallskip}
      4FGL J0506.9+0323 (76.73, 3.39) & $(2.72 \pm 0.37) \times 10^{-10}$ & 9.51  \\
      \noalign{\smallskip}
      4FGL J1012.3+0629 (153.08, 6.50) & $(1.19 \pm 0.06) \times 10^{-9}$ & 81.56  \\
      \noalign{\smallskip}
      4FGL J1016.0+0512 (154.01, 5.21) & $(9.42 \pm 0.51) \times 10^{-10}$ & 596.60  \\
      \noalign{\smallskip}
      4FGL J1018.4+0528 (154.62, 5.47) & $(3.23 \pm 0.41) \times 10^{-10}$ & 39.19  \\
      \noalign{\smallskip}
      4FGL J2118.0+0019 (319.50, 0.33) & $(1.67 \pm 0.34) \times 10^{-10}$ & 32.78  \\
      \noalign{\smallskip}
      4FGL J2223.3+0102 (335.85, 1.05) & $(1.46 \pm 0.31) \times 10^{-10}$ & 17.87  \\
      \noalign{\smallskip}
      4FGL J2226.8+0051 (336.71, 0.86) & $(3.38 \pm 0.53) \times 10^{-10}$ & 53.72  \\
      \noalign{\smallskip}
      4FGL J2227.9+0036 (336.98, 0.62) & $(9.08 \pm 0.67) \times 10^{-10}$ & 32.08  \\
      \noalign{\smallskip}
      4FGL J2252.6+1245 (343.17, 12.75) & $(1.49 \pm 0.31) \times 10^{-10}$ & 11.93  \\
      \noalign{\smallskip}
      \hline
   \end{tabular}
   
   \begin{flushleft}
     
   \end{flushleft}
\end{table}

\end{appendix}

\end{document}